   \documentclass[fleqn,usenatbib,usedcolumn]{mnras}
   \usepackage[british]{babel}             
   \usepackage{txfonts}                    
   \usepackage[T1]{fontenc}                
   \usepackage{graphicx}                   
    \hypersetup{pdfauthor={Timothy. A. Davis},
               pdftitle={The MASSIVE Survey -- III. Molecular gas and a broken Tully-Fisher relation in the most massive early-type galaxies},
               pdfkeywords={galaxies: elliptical and lenticular, cD -- ISM: molecules -- galaxies: ISM -- galaxies: evolution -- galaxies: star formation -- galaxies: kinematics and dynamics},
               bookmarksnumbered=true}
   \volume{{\rm in press}}

\usepackage{ae,aecompl}

\usepackage{amssymb}	
\usepackage{color,afterpage}
\usepackage{times,rotating}

\def\hi{\mbox{H\sc{i}}}

\def\atlas{{{ATLAS}}$^{\rm 3D}$}
\def\mum{$\mu$m}
\def\kms{km s$^{-1}$}

\def\msun{M$_{\odot}$}

\def\arcsec{$^{\prime \prime}$}
\definecolor{Mygrey}{gray}{0.75}

\newcommand{\gtsimeq}{\raisebox{-0.6ex}{$\,\stackrel{\raisebox{-.2ex}{$\textstyle >$}}{\sim}\,$}}
\newcommand{\farc}{\mbox{\ensuremath{.\!\!^{\prime\prime}}}}

\mathchardef\mhyphen="2D

\usepackage[compact]{titlesec}
\titlespacing{\section}{0pt}{*2}{*1}

\title[Molecular gas in MASSIVE galaxies]{The MASSIVE Survey -- III. Molecular gas and a broken Tully-Fisher relation in the most massive early-type galaxies} 
\author[Timothy A. Davis et al.]{\parbox{\textwidth}{Timothy A. Davis$^{1,2}$\thanks{E-mail: \texttt{DavisT@cardiff.ac.uk}}, Jenny Greene$^{3}$, Chung-Pei Ma$^{4}$, Viraj Pandya$^{3}$, John P. Blakeslee$^{5}$, Nicholas McConnell$^{6}$ and Jens Thomas$^{7}$}
\vspace{0.4cm}\\
\parbox{\textwidth}{$^{1}$Centre for Astrophysics Research, University of Hertfordshire, Hatfield, Herts AL10 9AB, UK\\
$^{2}$School of Physics \&\ Astronomy, Cardiff University, Queens Buildings, The Parade, Cardiff, CF24 3AA, UK\\
$^{3}$Department of Astrophysics, Princeton University, Princeton, NJ 08544, USA\\
$^{4}$Department of Astronomy, University of California, Berkeley, CA 94720, USA\\
$^{5}$Dominion Astrophysical Observatory, NRC Herzberg Institute of Astrophysics, Victoria, BC V9E 2E7, Canada\\
$^{6}$Institute for Astronomy, University of Hawaii at Manoa, Honolulu, HI 96822, USA\\
$^{7}$Max Planck-Institute for Extraterrestrial Physics, Giessenbachstr. 1, D-85741 Garching, Germany\\
}}
\begin{document}
\date{Accepted 2015 October 2.  Received 2015 October 2; in original form 2015 July 20}

\pagerange{\pageref{firstpage}--\pageref{lastpage}} \pubyear{2015}

\maketitle

\label{firstpage}

\begin{abstract}
In this work we present CO(1-0) and CO(2-1) observations of a pilot sample of 15 early-type galaxies (ETGs) drawn from the MASSIVE galaxy survey, a volume-limited integral-field spectroscopic study of the most massive ETGs ($M_* \gtsimeq10^{11.5}M_\odot$) within 108 Mpc. These objects were selected because they showed signs of an interstellar medium and/or star formation. A large amount of gas ($>$2$\times$10$^8$ \msun) is present in 10 out of 15 objects, and these galaxies have gas fractions higher than expected based on extrapolation from lower mass samples. We tentatively interpret this as evidence that stellar mass loss and hot halo cooling may be starting to play a role in fuelling the most massive galaxies. These MASSIVE ETGs seem to have lower star-formation efficiencies (SFE=SFR/M$_{\rm H2}$) than spiral galaxies, but the SFEs derived are consistent with being drawn from the same distribution found in other lower mass ETG samples. This suggests that the SFE is not simply a function of stellar mass, but that local, internal processes are more important for regulating star formation. Finally we used the CO line profiles to investigate the high-mass end of the Tully-Fisher relation (TFR). We find that there is a break in the slope of the TFR for ETGs at high masses (consistent with previous studies). The strength of this break correlates with the stellar velocity dispersion of the host galaxies, suggesting it is caused by additional baryonic mass being present in the centre of massive ETGs. We speculate on the root cause of this change and its implications for galaxy formation theories.
\end{abstract}


\section{Introduction}

More than half of the stars in the Universe today are found in elliptical galaxies, and yet their formation history remains mysterious, particularly at the high-mass end (e.g., \citealt{2007AJ....133.1741B}). 
A large range of open questions remain in our attempts to understand elliptical galaxy formation, including the connection between black hole accretion and galaxy growth \citep[e.g.][]{1998A&A...331L...1S,2001ApJ...554L.151B,2013ApJ...764..184M}, the variation in stellar initial mass function (IMF;  e.g. \citealt{2010Natur.468..940V,2010ApJ...709.1195T,2011MNRAS.415..545T,2012Natur.484..485C}), dark matter fraction \citep[e.g.][]{1993ApJ...416L..49R,2009MNRAS.396.1132T,2009ApJ...691..770T,2013MNRAS.432.1709C}, and the late-time assembly of galaxy outskirts \citep[e.g.][]{2007ApJ...658..710N,2009AJ....138.1417T,2015MNRAS.446..120D}. 
Another longstanding issue in understanding the evolution of the most massive galaxies is their lack of cold gas. Stars lose mass back into the interstellar medium (ISM), and the hot gas in massive halos usually has a short cooling time (e.g. \citealt{2014MNRAS.444..336C,2015arXiv150107111N}), yet the majority of these objects do not host large gas reservoirs. Evidence suggests that active galactic nuclei (AGN) prevent cooling flows in the most massive galaxy clusters \citep[e.g.][]{2012ARA&A..50..455F}, and thus suppress the formation of cold gas, but direct observational evidence for feedback in individual massive galaxies is sparse.

ATLAS$^{\rm 3D}$, a volume-limited survey of early-type galaxies (ETGs) within 40 Mpc, mapped out the demographics and kinematics of cold gas in massive galaxies (\citealt{2011MNRAS.414..940Y,2011MNRAS.414..968D,2011MNRAS.417..882D,2012MNRAS.421.1298C,2012MNRAS.422.1835S,2014MNRAS.444.3408Y}; see also \citealt{Welch:2003ev,Sage:2007jq,Welch:2010in}). These authors detected CO emission in 22\% of their sample objects, but showed that not all early-type galaxies are equally likely to harbour cold gas -- molecular gas appeared to be present only in galaxies with some net rotation while the most massive, non-rotating galaxies show no signs of molecular material. 
\hi\, on the other hand, was present in both galaxy types,  suggesting it is not always linked to the growth of an inner stellar disc \citep{2014MNRAS.444.3388S}.

In field environments evidence is mounting that the cold gas in ETGs has been acquired after the galaxy joined the red sequence, through {minor} merging or accretion of free-floating H{\small I} (e.g. \citealt{Sarzi:2006p1474,2011MNRAS.417..882D,2014MNRAS.444.3408Y}). In cluster environments such channels for acquiring cold gas are greatly suppressed as mergers are rare and free-floating gas easily destroyed. The same also seems to be true in the most massive ellipticals probed by \atlas, which host hot gas halos that can heat/destroy inflowing gas \citep{2011MNRAS.417..882D}. Despite this, the detection rate and the typical mass of H$_2$ found in ETGs does not vary as a function of mass, or across a range of cluster/field environments \citep{2011MNRAS.414..940Y}. 

In recent years it has become clear that even at fixed gas fraction, molecule-rich ETGs form stars less efficiently than normal spirals, and very much less efficiently than starburst galaxies \citep[see e.g.][]{2009ApJ...707..250M,2011MNRAS.415...61S,2012ApJ...758...73S,2013MNRAS.432.1914M,2014MNRAS.444.3427D,2015MNRAS.449.3503D}. As part of \atlas, \cite{2014MNRAS.444.3427D} found that the high-mass ETGs with relaxed molecular gas reservoirs deeply embedded in the potential well were most likely to show signs of a low star formation efficiency (SFE=SFR/M$_{\rm H2}$). 

In all, \atlas\ made important strides in advancing our understanding of the cold gas phase in ETGs. However, this survey was limited by the small volume probed, and did not contain many of the most massive, slowly rotating elliptical galaxies. 
We aim to address this problem with MASSIVE, a survey of the  $\approx$100 most massive galaxies within 108 Mpc, using a combination of wide-field and adaptive-optics-assisted integral field spectroscopy (IFS). This volume-limited survey targets a distinct stellar mass range ($M_* \gtsimeq10^{11.5}M_\odot$; $M_K{\,<\,}{-}25.3$ mag) that has not been systematically studied to date. Full details of this survey are presented in \cite{2014ApJ...795..158M}. In this work, Paper III of the MASSIVE survey, we present a pilot study attempting to detect molecular gas in these high mass galaxies. We describe IRAM-30m observations of a sample of 15 MASSIVE galaxies, and study their gas masses, star-formation efficiencies and circular velocities. In a future work this pilot survey will be extended to give a full picture of the gas content in very massive ETGs. 

In Section \ref{sample} of this paper we present details of our sample selection and the properties of the target objects. Section \ref{data} details the observation parameters and reduction. In Section \ref{results} we present our results, before concluding in Section \ref{conclude}. 

\section{Sample}
\label{sample}

 \begin{table*}
\caption{Properties of the MASSIVE ETGs included in this pilot study}
\begin{tabular*}{0.85\textwidth}{@{\extracolsep{\fill}}l r r r r r r r r}
\hline
Name & Distance & inc & M$_{Ks}$ & $\sigma$ & log$_{10}$(L$_{\rm 22,corr}$) & SFR & Selection & Good inc? \\ 
  & (Mpc) & ($^\circ$) & (mag) & (\kms) & (ergs s$^{-1}$) & (M$_{\odot}$ yr$^{-1}$) &  & \\
 (1) & (2) & (3) & (4) & (5) & (6) & (7) & (8) & (9)\\
\hline
IC0310 &       77.5 &         27$\pm$         5 &     -25.35 &       230. &      42.27$\pm$      0.05 &       0.33$\pm$      0.03 & 22\mum & -\\
NGC0665 &       74.6 &         34$\pm$         5 &     -25.51 &       190. &      42.21$\pm$      0.05 &       0.29$\pm$      0.03 & 22\mum & $\checkmark$\\
NGC0708 &       69.0 &         42$\pm$         5 &     -25.65 &       230. & $<$     41.89 & $<$      0.15 & Lit. & -\\
NGC0997 &       90.4 &         34$\pm$         5 &     -25.40 &       260. &      42.45$\pm$      0.05 &       0.47$\pm$      0.04 & Dust & $\checkmark$\\
NGC1132 &       97.6 &         45$\pm$         5 &     -25.70 &       246. &      41.87$\pm$      0.05 &       0.14$\pm$      0.01 & 22\mum & -\\
NGC1167 &       70.2 &         29$\pm$         5 &     -25.64 &       204. &      42.21$\pm$      0.05 &       0.29$\pm$      0.03 & Lit. & $\checkmark$\\
NGC1497 &       87.8 &         85$\pm$         5 &     -25.31 &       249. &      42.67$\pm$      0.05 &       0.74$\pm$      0.07 & Dust & -\\
NGC1684 &       63.5 &         42$\pm$         5 &     -25.34 &       306. &      42.29$\pm$      0.05 &       0.34$\pm$      0.03 & 22\mum & -\\
NGC2258 &       59.0 &         37$\pm$         5 &     -25.66 &       287. &      41.49$\pm$      0.05 &       0.07$\pm$      0.01 & Dust & -\\
NGC2320 &       89.4 &         60$\pm$         7 &     -25.93 &       315. &      42.18$\pm$      0.05 &       0.27$\pm$      0.02 & Lit. & $\checkmark$\\
NGC5208 &      105.0 &         69$\pm$        14 &     -25.61 &       252. &      42.56$\pm$      0.05 &       0.58$\pm$      0.05 & Dust & $\checkmark$\\
NGC5252 &      103.8 &         64$\pm$         8 &     -25.32 &       196. &      43.34$\pm$      0.05 &       2.86$\pm$      0.25 & 22\mum & $\checkmark$\\
NGC6482 &       61.4 &         42$\pm$         5 &     -25.60 &       322. & $<$     41.94 & $<$      0.17 & Dust & -\\
NGC7052 &       69.3 &         72$\pm$         5 &     -25.67 &       284. &      42.15$\pm$      0.05 &       0.26$\pm$      0.02 & Lit. & $\checkmark$\\
NGC7556 &      103.0 &         44$\pm$         5 &     -25.83 &       268. &      42.71$\pm$      0.05 &       0.81$\pm$      0.07 & 22\mum & -\\
        \hline
\end{tabular*}
\parbox[t]{0.85\textwidth}{ \textit{Notes:}  Column 1 lists the name of each source. Column 2, 4 and 5 are the distance, $Ks$-band absolute magnitude and velocity dispersion, all reproduced here from \cite{2014ApJ...795..158M}. Column 3 contains the inclination, estimated as described in Section \ref{inc_est}. Column 6 shows the 22 \mum\ luminosity of this source, after correction for the old stellar contribution as described in Section \ref{sfr_describe}. Column 7 shows the star formation rate derived from this luminosity, using the calibration of \cite{2007ApJ...666..870C}. Column 8 lists the selection method used to include this galaxy in our pilot sample (see Section \ref{sample}). Column 9 shows objects where we were able to determine an inclination from ellipse fitting to dust structures or by modelling resolved observations (see Section \ref{inc_est}).}
\label{proptable}
\end{table*}

In this work we study a sample of 15 galaxies from the MASSIVE survey \citep{2014ApJ...795..158M}. We present new observations of 11 ETGs, and include 4 detections from literature.
MASSIVE is a volume-limited, multi-wavelength,
spectroscopic and photometric survey of the most massive galaxies in the
local universe.  The full sample includes 116 galaxies in the northern
sky with distance $D < 108$ Mpc and absolute $K$-band magnitude
$M_K{\,<\,}{-}25.3$ mag, corresponding to stellar masses M$_* \gtsimeq
10^{11.5}M_\odot$.  
The MASSIVE survey volume is more than an order of magnitude larger than that probed by \atlas\ and there are only 6 overlapping galaxies in the two surveys.
Wide-field IFS data, taken using the Mitchell Spectrograph (formerly called
VIRUS-P; \citealt{2008SPIE.7014E..06H}) at McDonald Observatory, are being obtained for all objects in the MASSIVE survey, enabling us to derive stellar and gas parameters (and study the galaxies stellar populations; \citealt{2015arXiv150402483G}) out to beyond $\sim$2 effective radii. 

The objects in this paper form a pilot sample, selected to either have optically obscuring dust visible in \emph{Hubble Space Telescope} imaging (which often goes hand in hand with the presence of a cold ISM in ETGs; \citealt{2013MNRAS.432.1796A,2015MNRAS.449.3503D}), or to have strong 22 \mum\ emission, over and above that expected from circumstellar dust around old stars (see \citealt{2014MNRAS.444.3427D}). 
We add to this sample 4 MASSIVE galaxies which have existing CO detections in the literature \citep{2010A&A...518A...9O,2015A&A...573A.111O,2005ApJ...634..258Y}. 
Table \ref{proptable} contains the properties of all these objects, and details the method used to select each object.

 Selecting on these proxies for the presence of an ISM/star formation means that this pilot sample is highly biased towards gas-rich objects, and statistical comparison of these detections with other samples is difficult. Despite this bias, with respect to some other galaxy properties these objects are reasonably representative of the parent sample. We do not sample the most massive objects, but these are also relatively rare. Our objects all have $M_K{\,>\,}{-}25.95$ mag while NGC4889, the highest mass galaxy in MASSIVE, has $M_K$=${-}26.64$ mag. Kolmogorov-Smirnov and Mann-Whitney U tests are unable to reject the null hypothesis that these objects are drawn randomly from the parent MASSIVE sample in terms of mass and velocity dispersion. 

This subsample of objects does, however, seem to be somewhat biased in terms of shape. 
Over half of the sample objects have ellipticities greater than 0.3, which is somewhat larger than the fraction of such flattened objects in the sample as a whole \citep{2014ApJ...795..158M}.
This may be evidence that our selection criteria lead us to select the intrinsically flatter, fast rotating galaxies which have central power-law cusps in their density profile, over the giant, cored (possibly triaxial) slow rotators. Full exploration of this possibility is left to a future work, once data is available to study the properties of these objects in detail.

\subsection{Inclinations}
\label{inc_est}
In later sections of this paper we require an estimate of the intrinsic inclination of the galaxy, in order to de-project the observed CO velocity widths and study the Tully-Fisher relation (TFR) of ETGs. \cite{2011MNRAS.414..968D} studied in some detail different methods for estimating inclinations in ETGs and the effect these have on the derived TFR. Well-resolved kinematic observations of the gas itself provide the best inclination estimates. In the absence of such data the morphology of dust lanes or the observed axial ratio of the galaxy can be used to estimate the inclination. These proxies add scatter to the derived TFR, but do not bias its best fit slope or intercept. 

 In these objects we estimate the inclination using interferometric imaging, the dust-lane morphology, or the galaxy axial ratio where the other methods are not possible. NGC2320 is the only sample galaxy which has been observed using aperture synthesis techniques, providing an interferometric map of its CO distribution which has been modelled in \cite{2005ApJ...634..258Y}. We use their model estimate of the galaxy inclination here. 
 
 Six of our objects have clear, regular dust lanes visible in optical imaging. In these objects the dust lanes appear to have a flat, intrinsically circular geometry, and thus ellipse fitting to the image yields an inclination directly from the axial ratio of the fitted ellipse. We do not try to estimate inclinations from objects where the dust is filamentary or patchy. Following \cite{2011MNRAS.414..968D} we consider dust and interferometric inclinations to be relatively robust, and indicate the objects with either of these inclination estimates available with a tick in Table \ref{proptable}.
  
Where no dust lane or interferometric observations of our sources exist we estimate the galaxy inclination using the axial ratios tabulated in the NASA/IPAC Extragalactic Database (NED). These axial ratios cannot be simply converted into inclinations due to the intrinsic shapes of early-type objects, which have large bulges and thus a significant thickness. Following \cite{2011MNRAS.414..968D} we correct the observed axis ratios using the following relation
 
 \begin{equation}
\label{comaxial}
i_{\rm b/a} = \cos^{-1}\left(\sqrt{\frac{q^2-q_0^2}{1-q_0^2}}\,\right),
\end{equation}
\noindent where $q$ is the ratio of the semi-minor to the semi-major axis of the galaxy, and $q_0$ is the intrinsic axial ratio when the galaxy is seen edge-on. Various lines of enquiry suggest a mean $q_0$ value of $\approx$0.34 for the objects which are CO detected   in the \atlas\ sample \citep{2014MNRAS.444.3340W}. We use this axial ratio in this work. We note that this value is only likely to be applicable if these objects are scaled up versions of the \atlas\ fast rotators (see \citealt{2014MNRAS.444.3340W}), however if these objects are intrinsically rounder (like the \atlas\ slow rotators) this would bias our inclinations towards low values. The inclination errors quoted in Table \ref{proptable} do not take this intrinsic shape uncertainty into account, and are simply estimated by propagating forward the reported error in the axial ratio, with a minimum assumed error of 5 degrees. However, even in the most extreme case, where all of the objects without good inclinations are actually observed edge-on, this does not change the conclusions of this paper. 

\subsection{Star-formation rates}
\label{sfr_describe}
In order to estimate the star-formation rate (SFR) in these objects, we make use of \textit{WISE} \citep{2010AJ....140.1868W} 22 \mum\ data. 
Emission at $\approx$20-25 \mum\ traces warm dust, that is present around hot newly-formed stars, in the ejected circumstellar material around hot old stars, and in AGN torii. If one can correct for the emission from old stars (in the absence of strong AGN), the $\approx$20-25 \mum\ emission can provide a sensitive estimate of the amount of obscured star formation in our systems.

Here we use 22~\mum\ catalogue aperture magnitude values (parameter \textit{w4gmag}) from the \textit{WISE} catalogue \citep{2010AJ....140.1868W} all sky data release. The aperture values are calculated using elliptical apertures defined from the position, size and inclination of the galaxy from the Two Micron All Sky Survey \citep[2MASS][]{Skrutskie:2006p2829} Extended Source Catalog (XSC; \citealt{Jarrett:2000p2407}), and enlarged by the \textit{WISE} team to correct for the larger point-spread function of the \textit{WISE} satellite. See the \textit{WISE} documentation\footnote{http://wise2.ipac.caltech.edu/docs/release/allsky/ - accessed 20/05/15} for full details of these magnitudes. 

We correct the observed fluxes, subtracting the flux contributed from circumstellar emission around old stars using the prescription described in detail in \cite{2014MNRAS.444.3427D}. 
The corrected 22~\mum\ luminosities we measure for each object (and the respective errors) are listed in Table \ref{proptable} for our CO detected sample. Two objects had fluxes low enough that we were unable to distinguish a star formation related component from that expected from old stars, and these are listed as upper limits in the table. We use the prescription of \citet[which was benchmarked against 24\mum\ SFRs in ETGs in \citealt{2014MNRAS.444.3427D}]{2007ApJ...666..870C} to convert these 22 \mum\ luminosities into SFRs, and again list these in Table \ref{proptable}. This prescription includes a correction factor that attempts to account for the unobscured star formation not visible in the infrared. If our objects have larger fractions of unobscured star formation than typical spirals then this could cause us to underestimate the SFR. This was not found to be the case for the ETGs in the study of \cite{2014MNRAS.444.3427D}, however, and so we do not expect this assumption to bias our results.

\section{IRAM-30m observations}
\label{data}

    \begin{figure*}
\begin{minipage}[t]{1.0\textwidth}
\begin{center}
$\begin{array}{cccc}
\begin{turn}{90}\large \hspace{1.1cm} IC0310\end{turn} &
\includegraphics[width=5cm,angle=0,clip,trim=0.0cm 1.3cm 0cm 0.0cm]{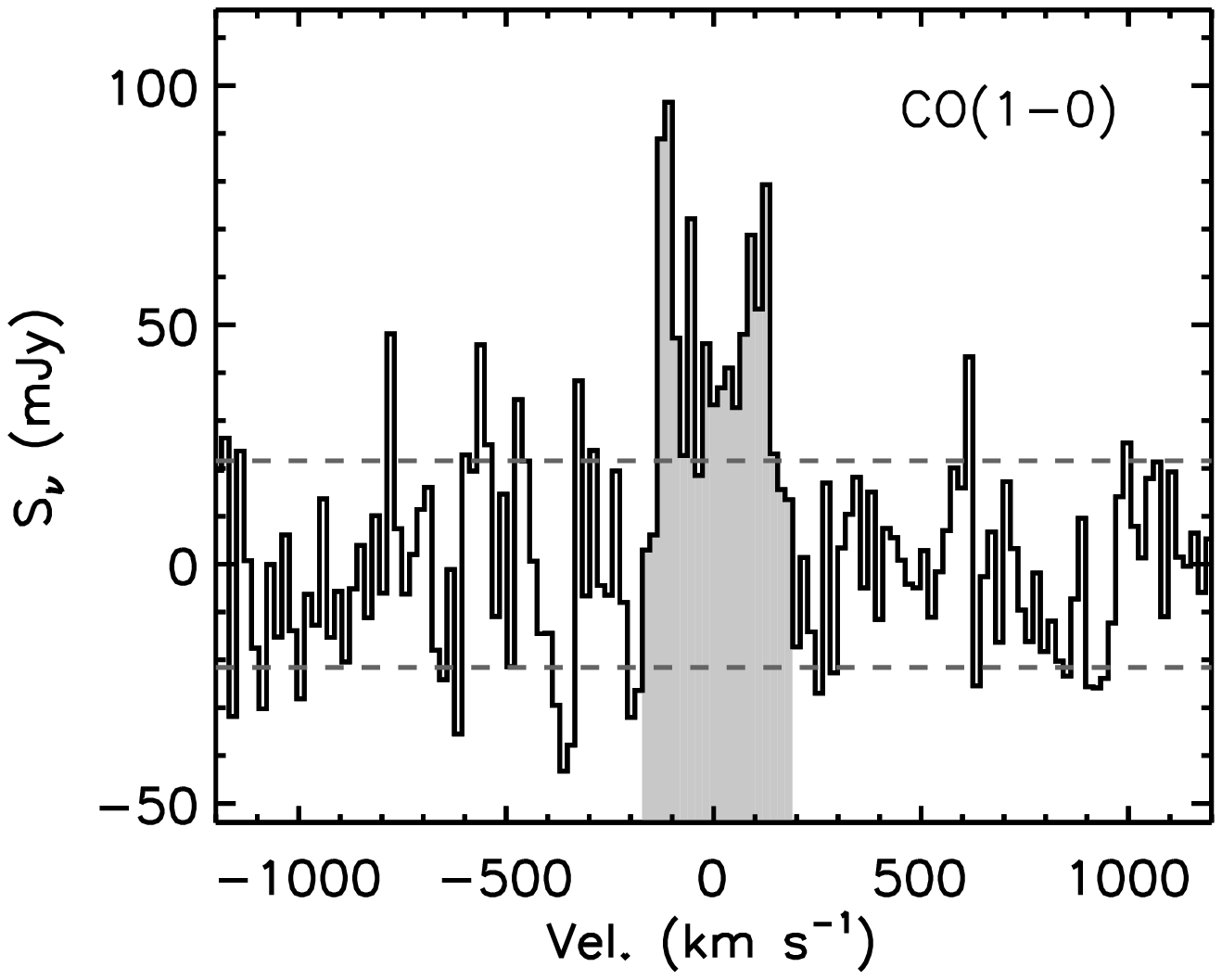} &
\includegraphics[width=5cm,angle=0,clip,trim=0.0cm 1.3cm 0cm 0.0cm]{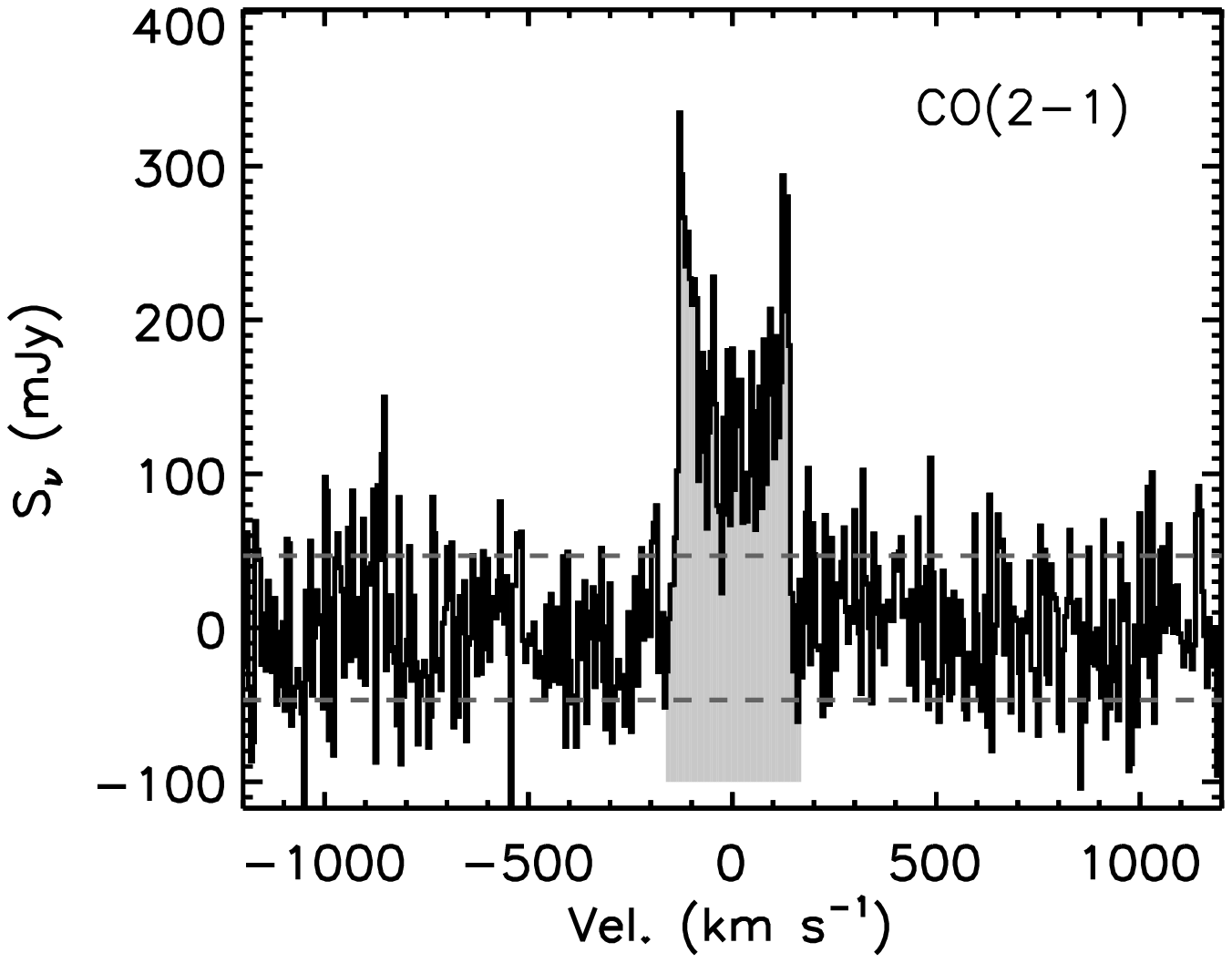} &
\includegraphics[width=4.7cm,angle=0,clip,trim=0.0cm 1.3cm 0cm 0.0cm]{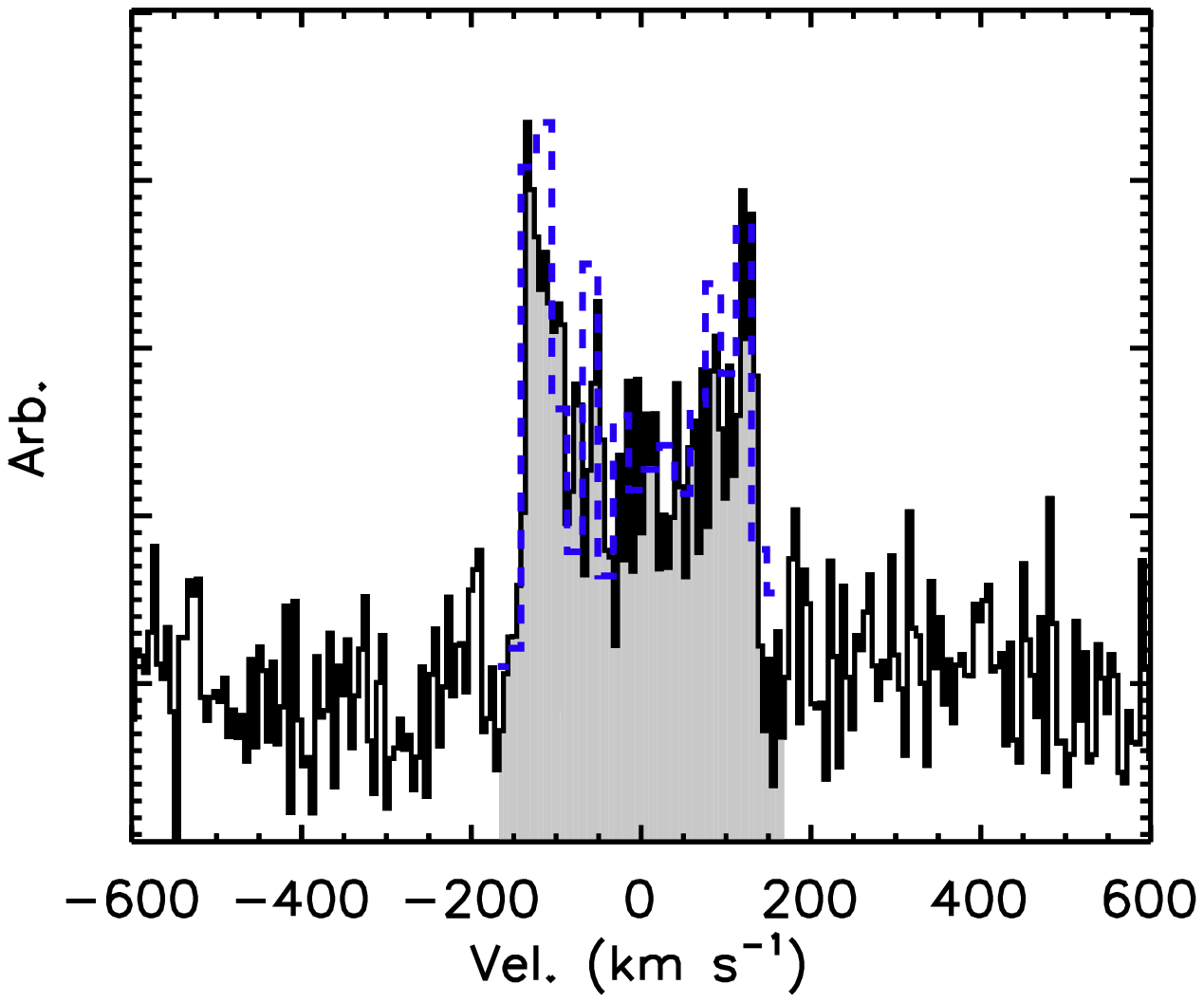} \\
\begin{turn}{90}\large \hspace{1.1cm} NGC0665\end{turn} &
\includegraphics[width=5cm,angle=0,clip,trim=0.0cm 1.3cm 0cm 0.0cm]{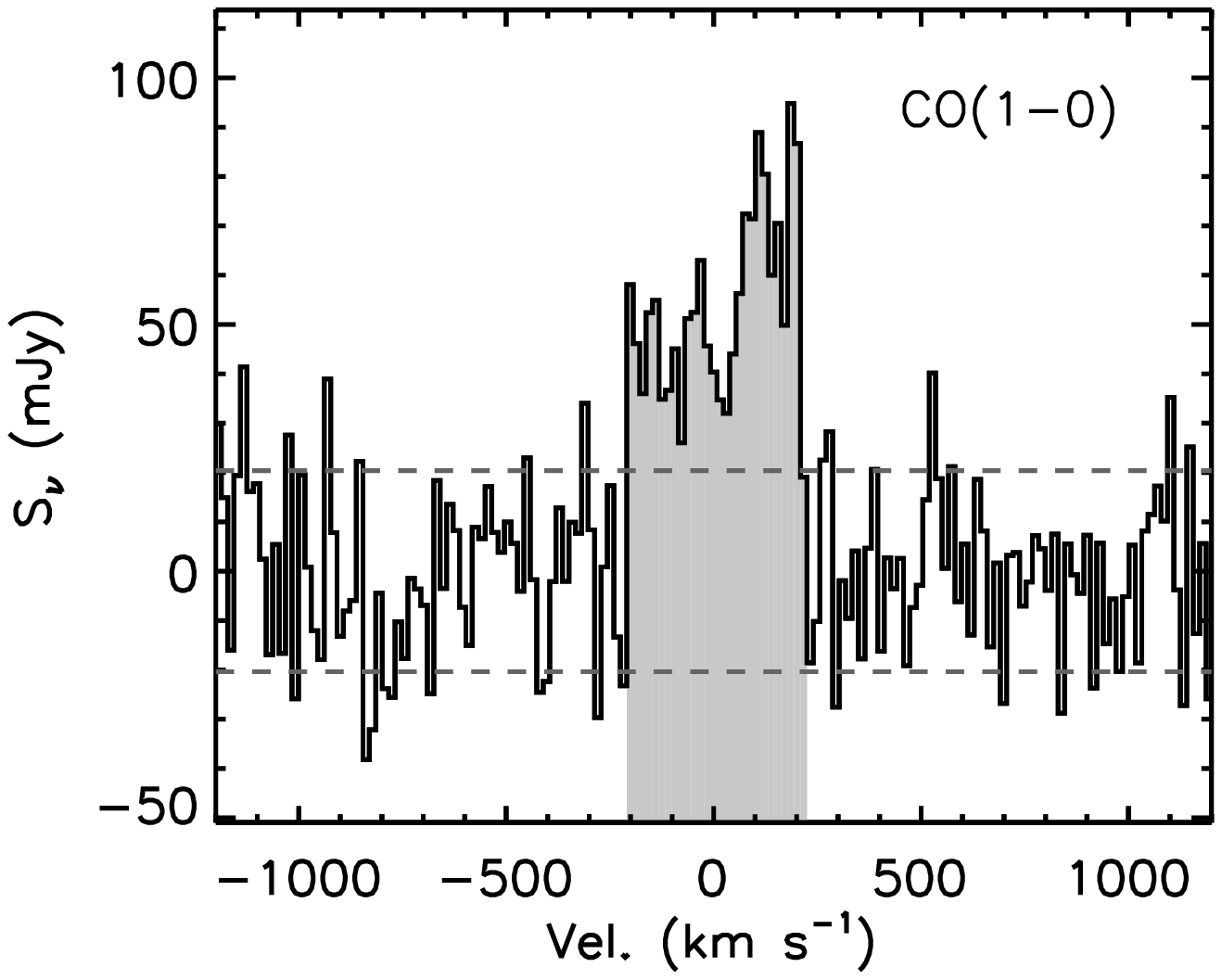} &
\includegraphics[width=5cm,angle=0,clip,trim=0.0cm 1.3cm 0cm 0.0cm]{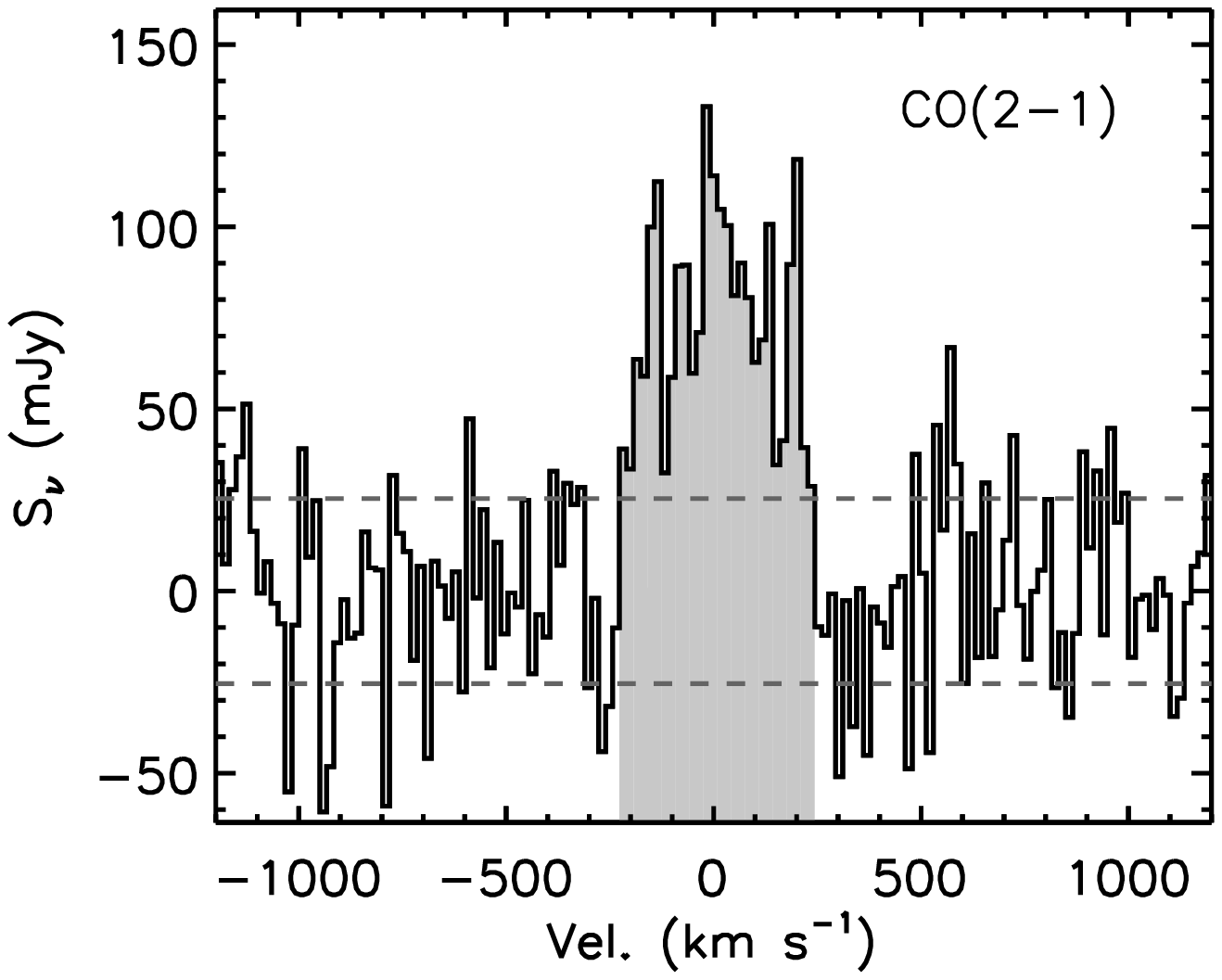} &
\includegraphics[width=4.7cm,angle=0,clip,trim=0.0cm 1.3cm 0cm 0.0cm]{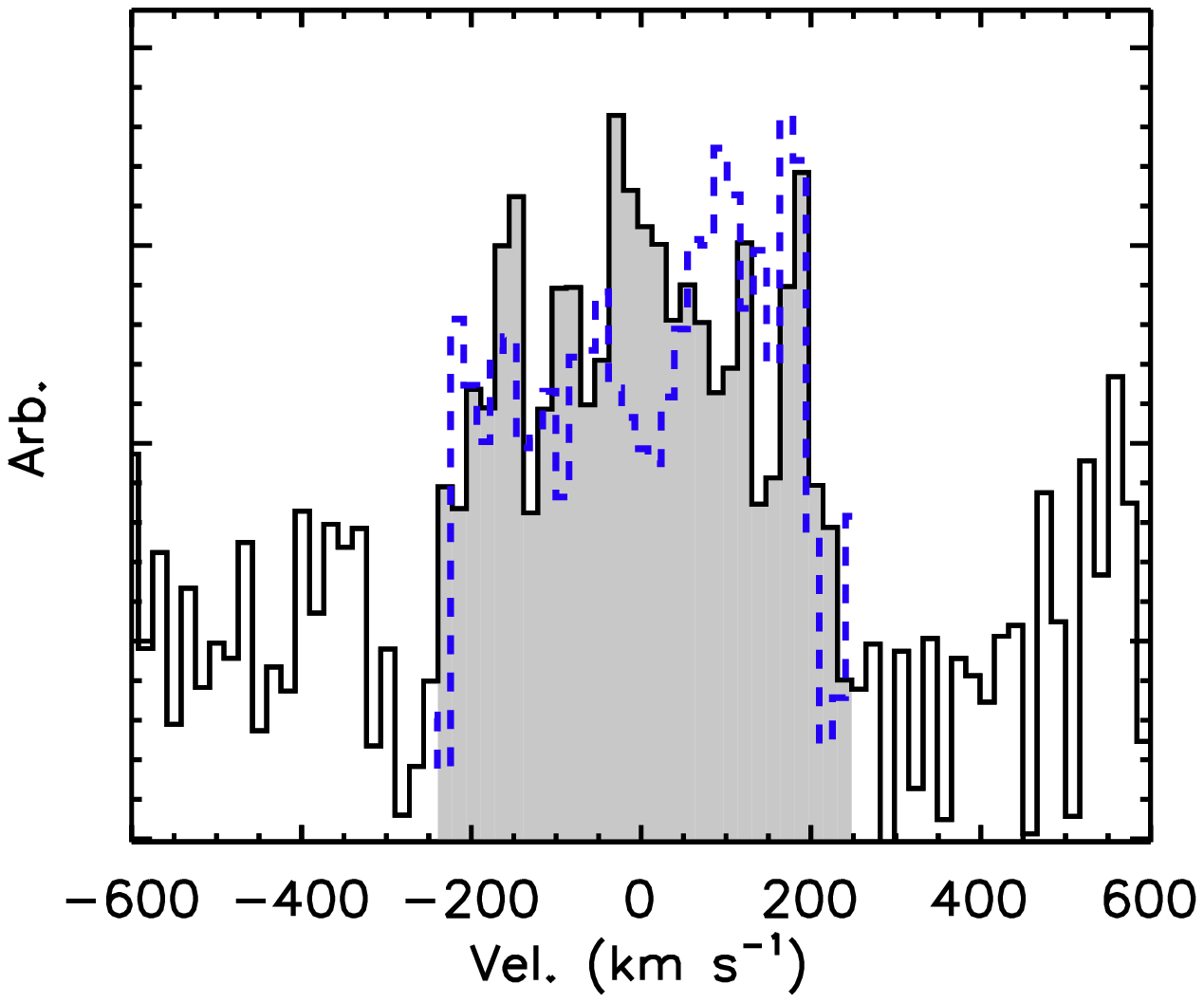} \\
\begin{turn}{90}\large \hspace{1.1cm} NGC0997 \end{turn} &
\includegraphics[width=5cm,angle=0,clip,trim=0.0cm 1.3cm 0cm 0.0cm]{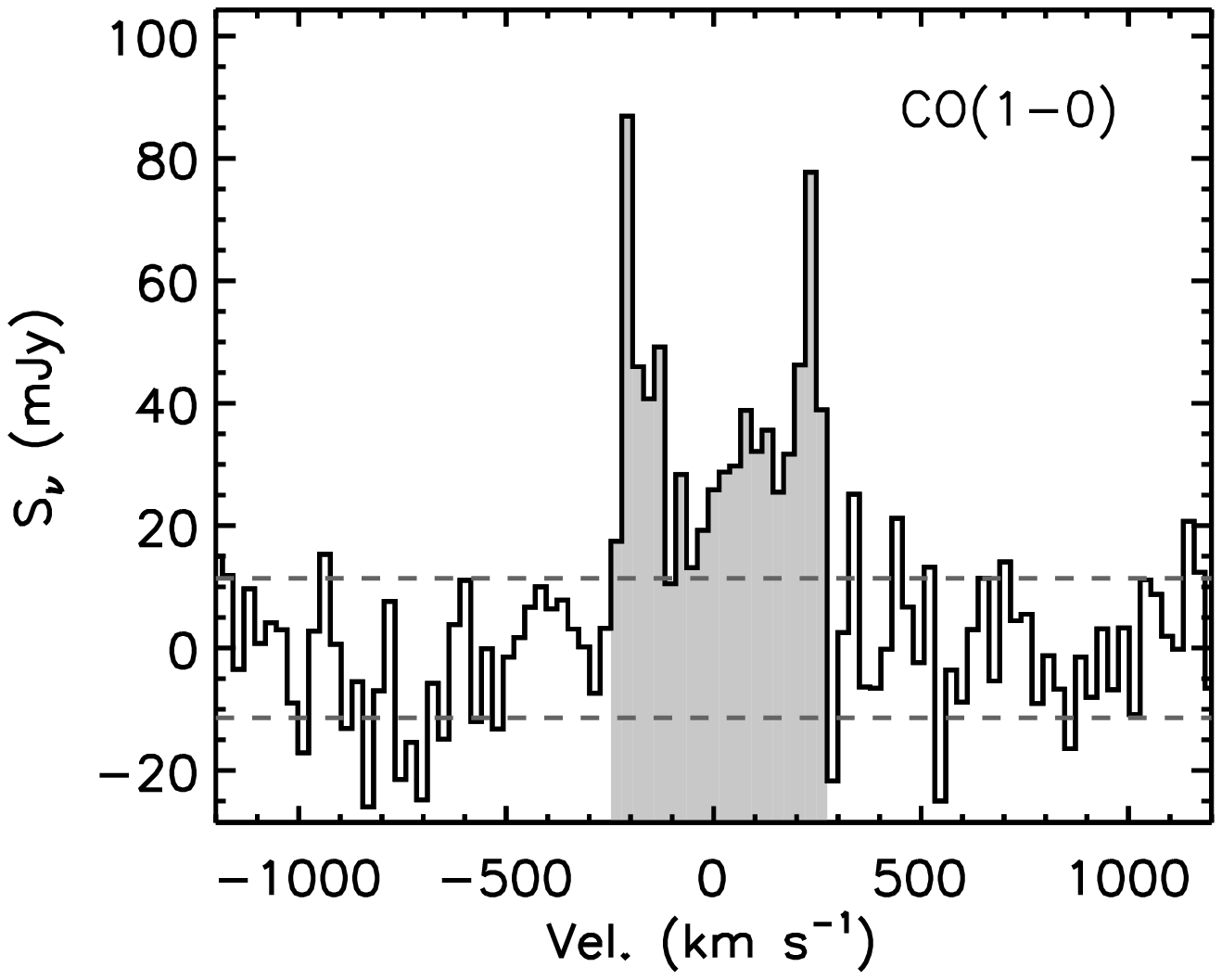} &
\includegraphics[width=5cm,angle=0,clip,trim=0.0cm 1.3cm 0cm 0.0cm]{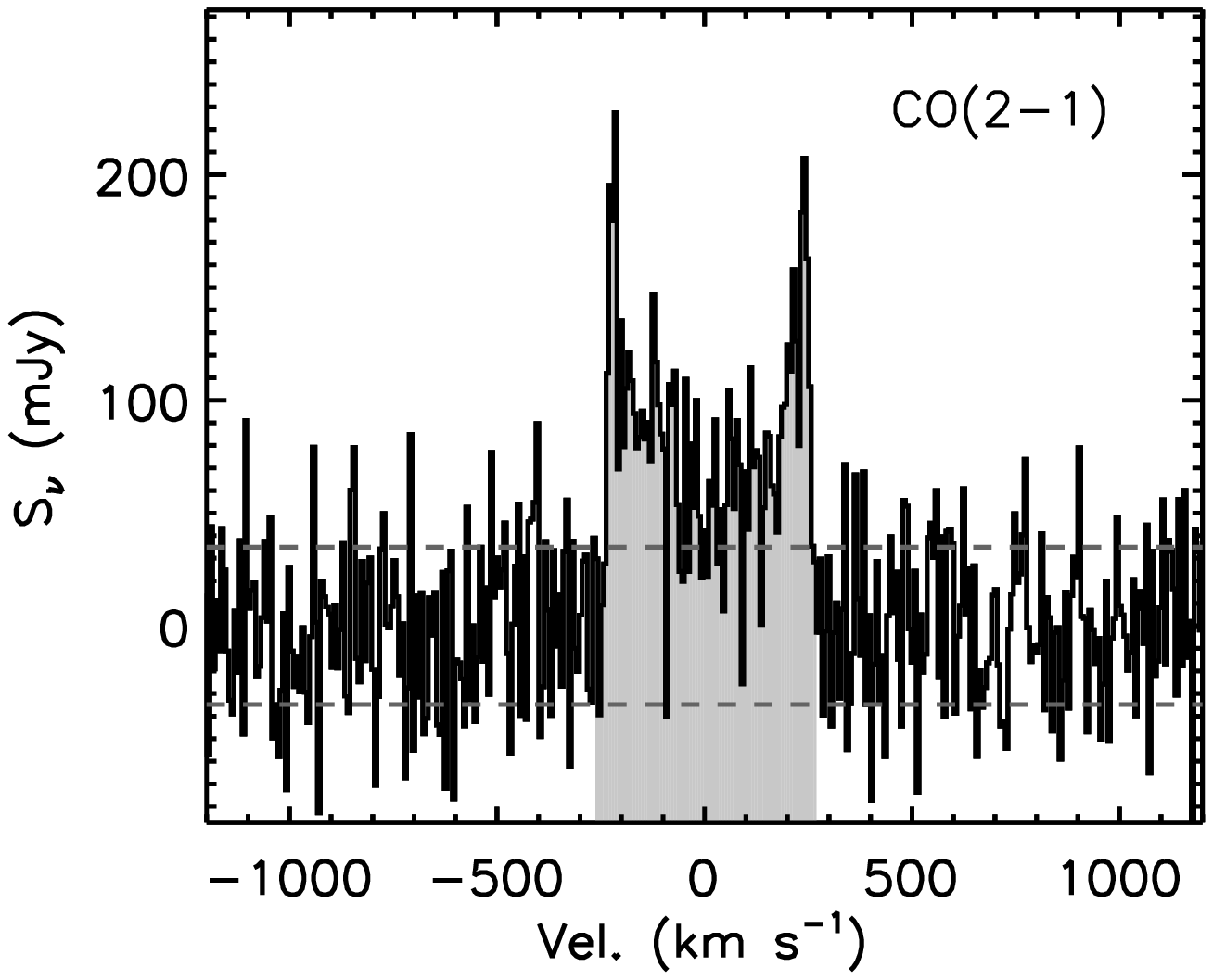} & 
\includegraphics[width=4.7cm,angle=0,clip,trim=0.0cm 1.3cm 0cm 0.0cm]{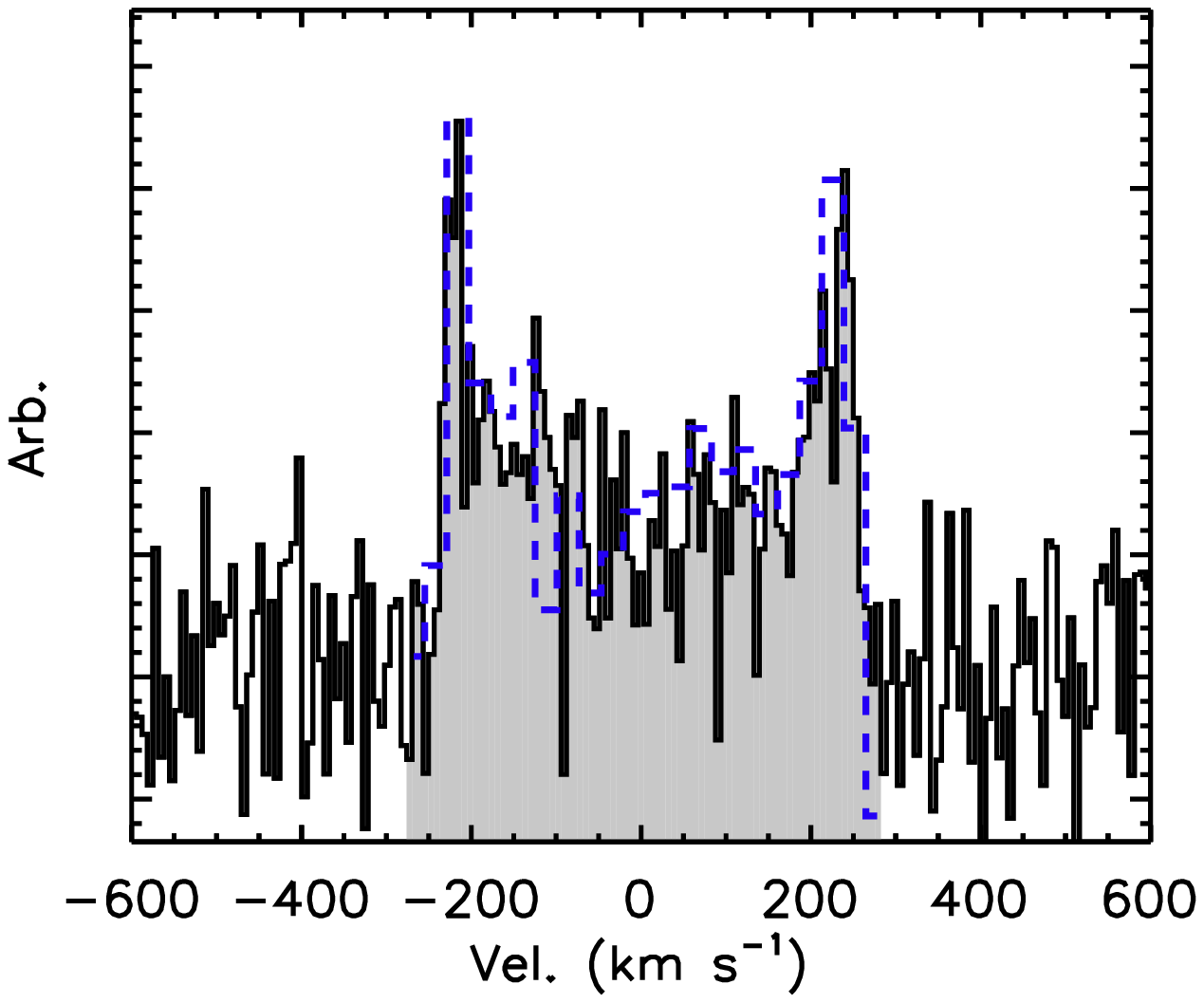} \\
\begin{turn}{90}\large \hspace{1.1cm} NGC1497 \end{turn} &
\includegraphics[width=5cm,angle=0,clip,trim=0.0cm 1.3cm 0cm 0.0cm]{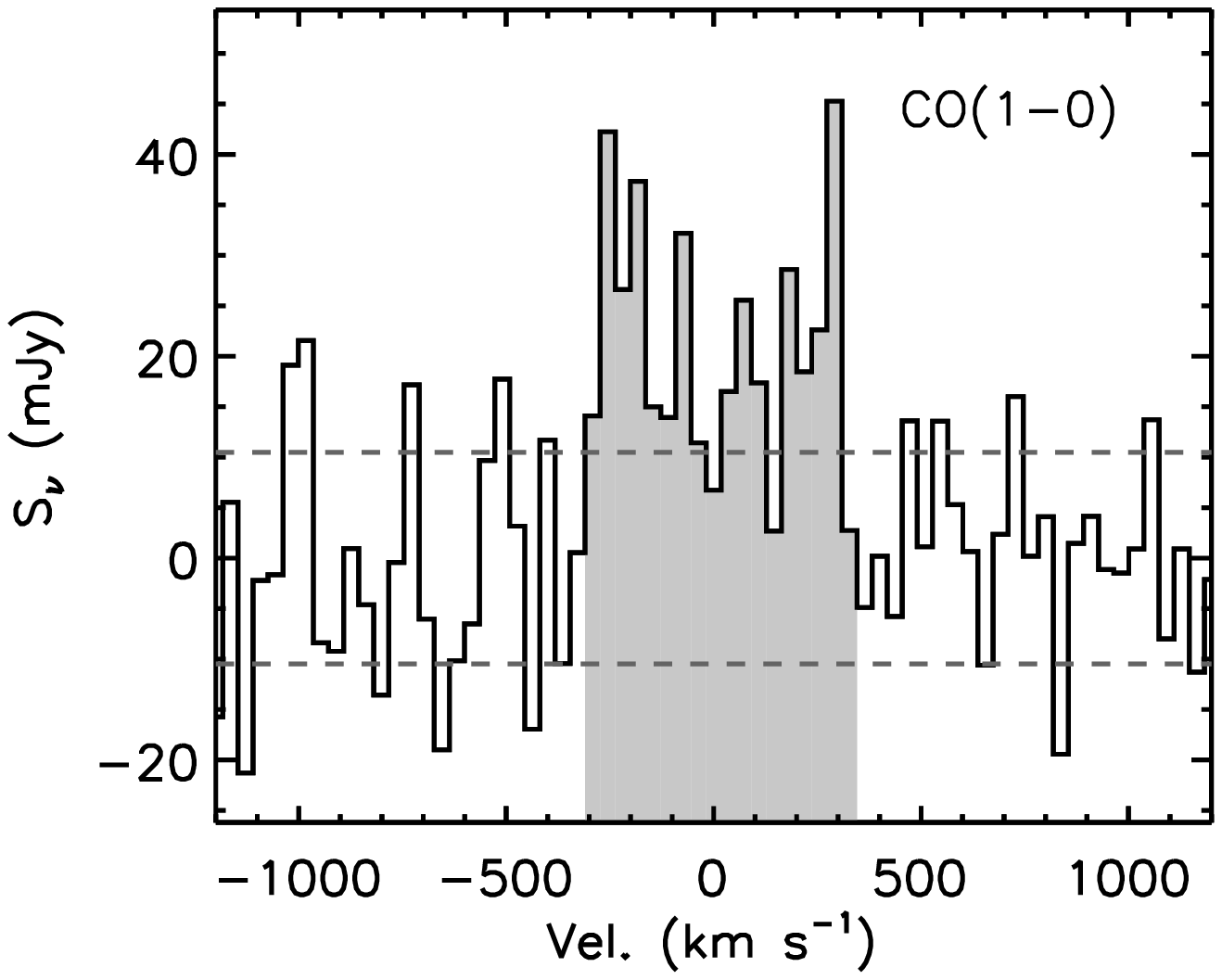} &
\includegraphics[width=5cm,angle=0,clip,trim=0.0cm 1.3cm 0cm 0.0cm]{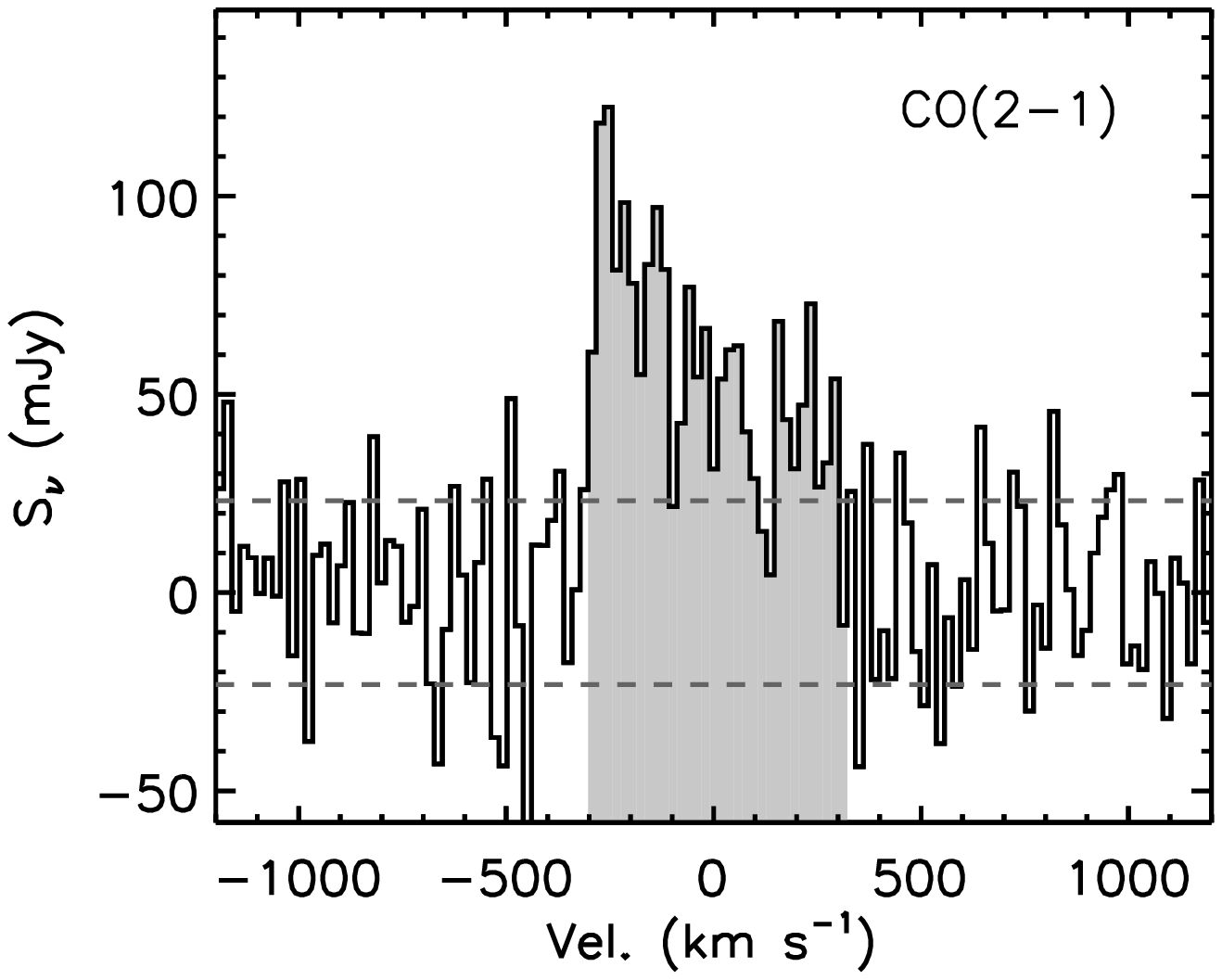}& 
\includegraphics[width=4.7cm,angle=0,clip,trim=0.0cm 1.3cm 0cm 0.0cm]{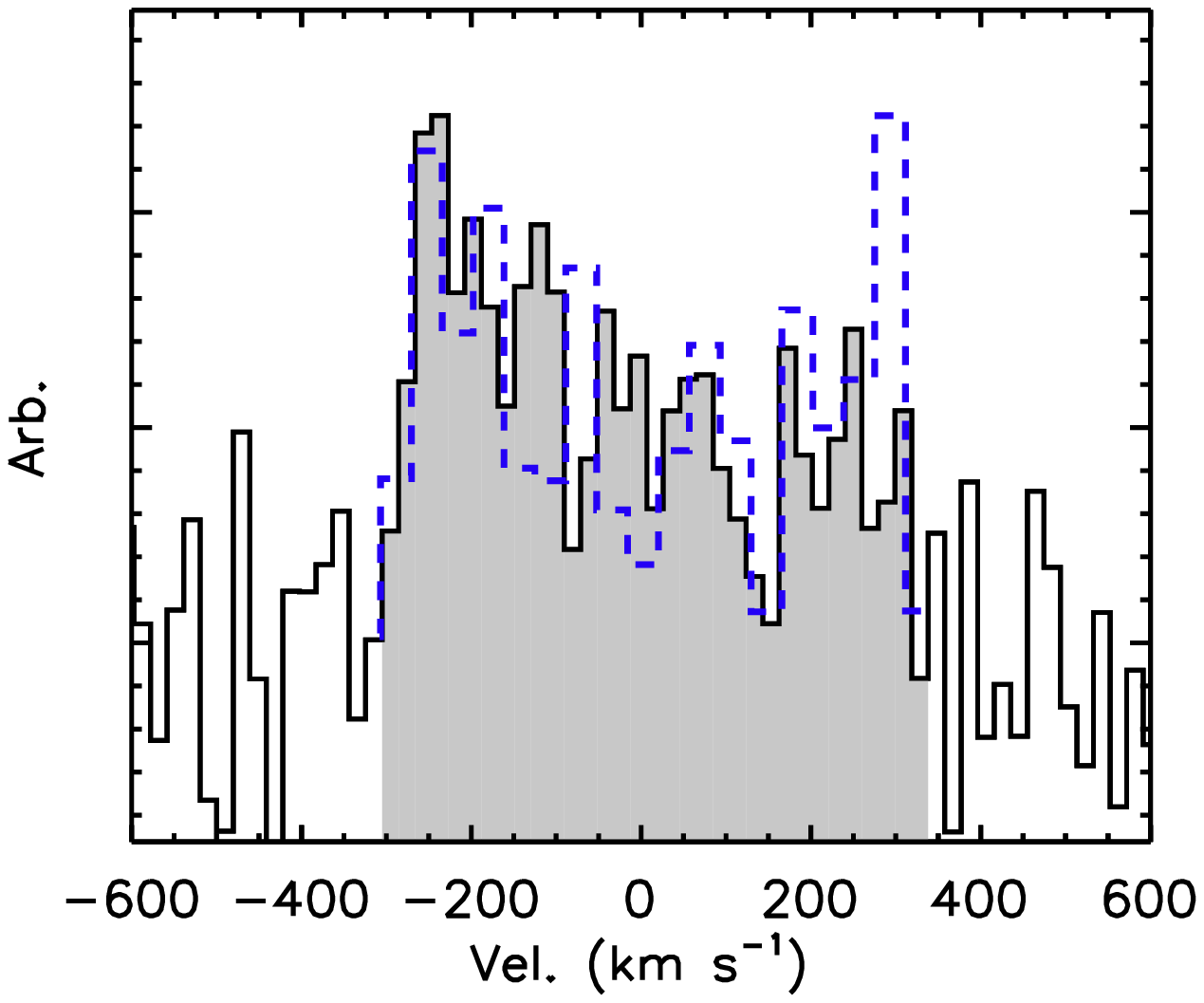} \\
\begin{turn}{90}\large \hspace{1.1cm} NGC1684\end{turn} &
\includegraphics[width=5cm,angle=0,clip,trim=0.0cm 1.3cm 0cm 0.0cm]{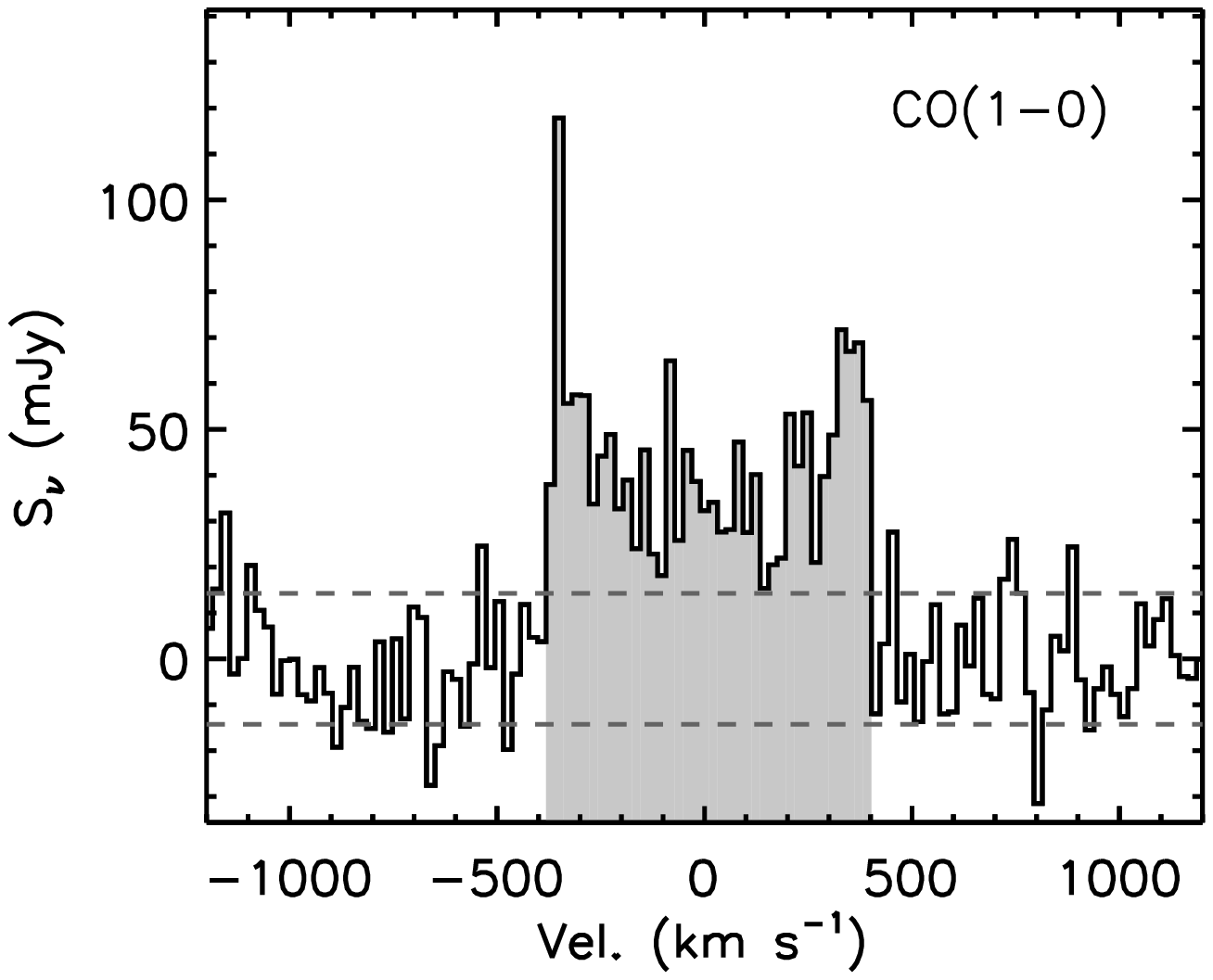} &
\includegraphics[width=5cm,angle=0,clip,trim=0.0cm 1.3cm 0cm 0.0cm]{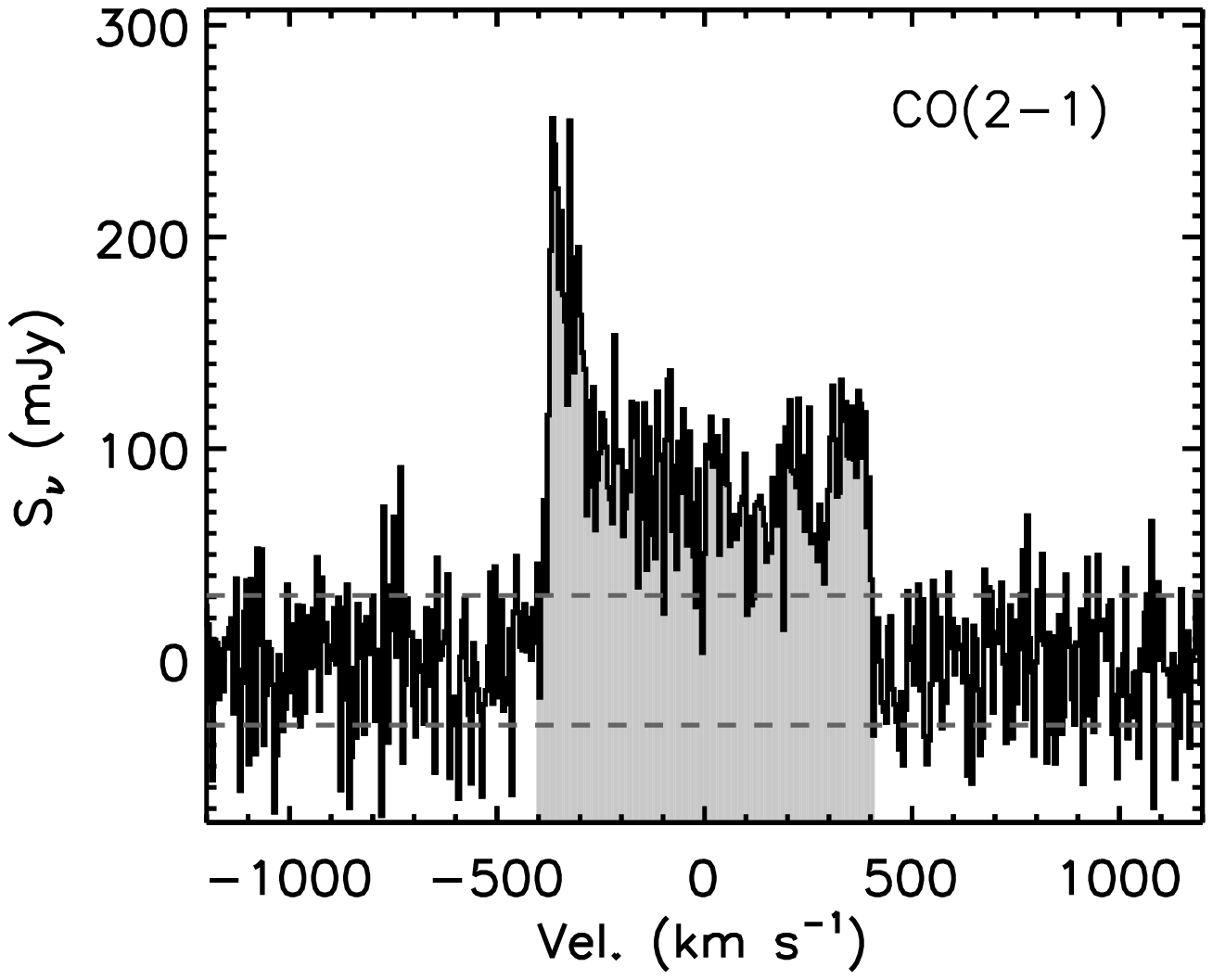} &
\includegraphics[width=4.7cm,angle=0,clip,trim=0.0cm 1.3cm 0cm 0.0cm]{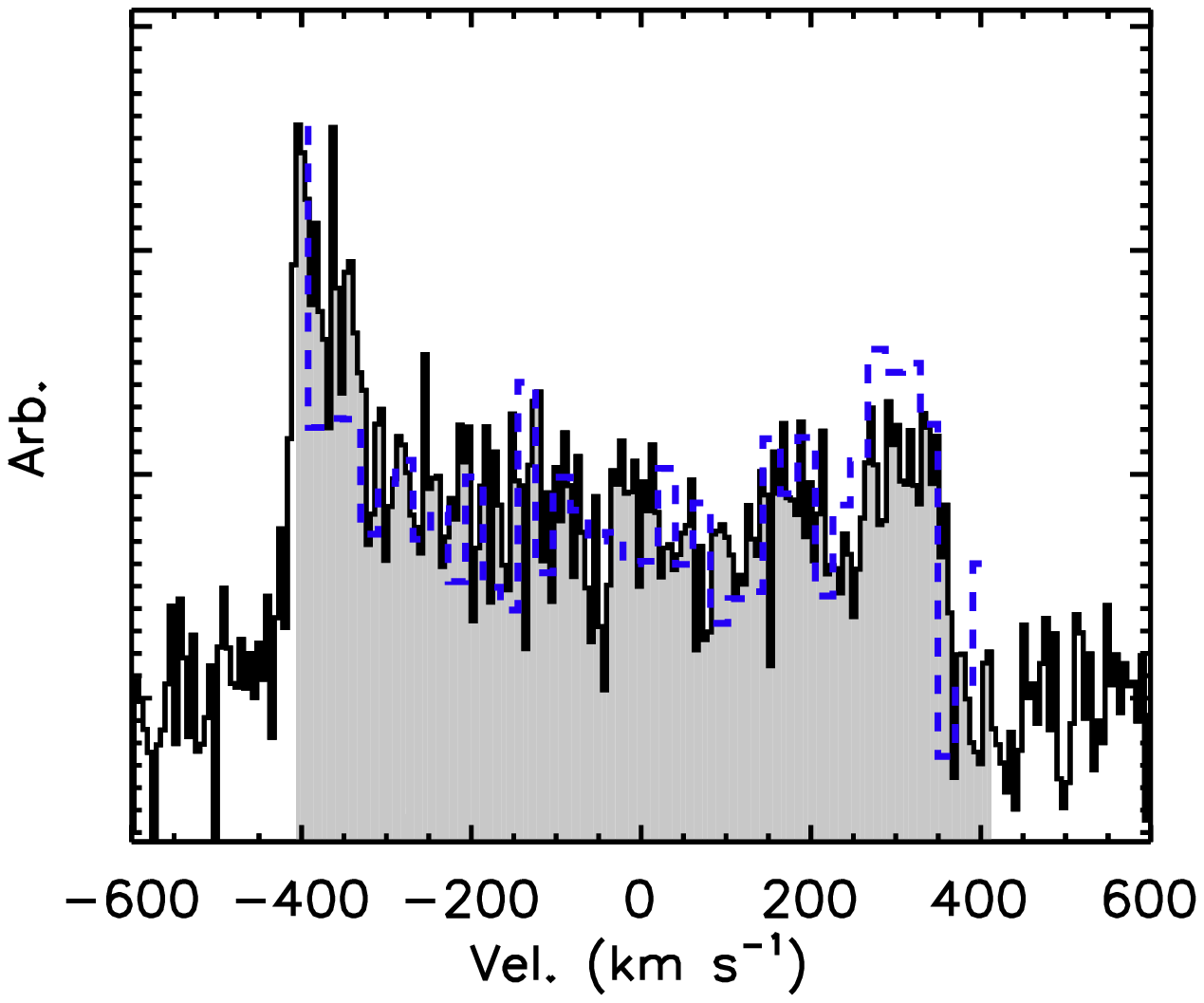} \\
\begin{turn}{90}\large \hspace{1.1cm} NGC5208 \end{turn} &
\includegraphics[width=5cm,angle=0,clip,trim=-0.2cm 0cm 0cm 0.0cm]{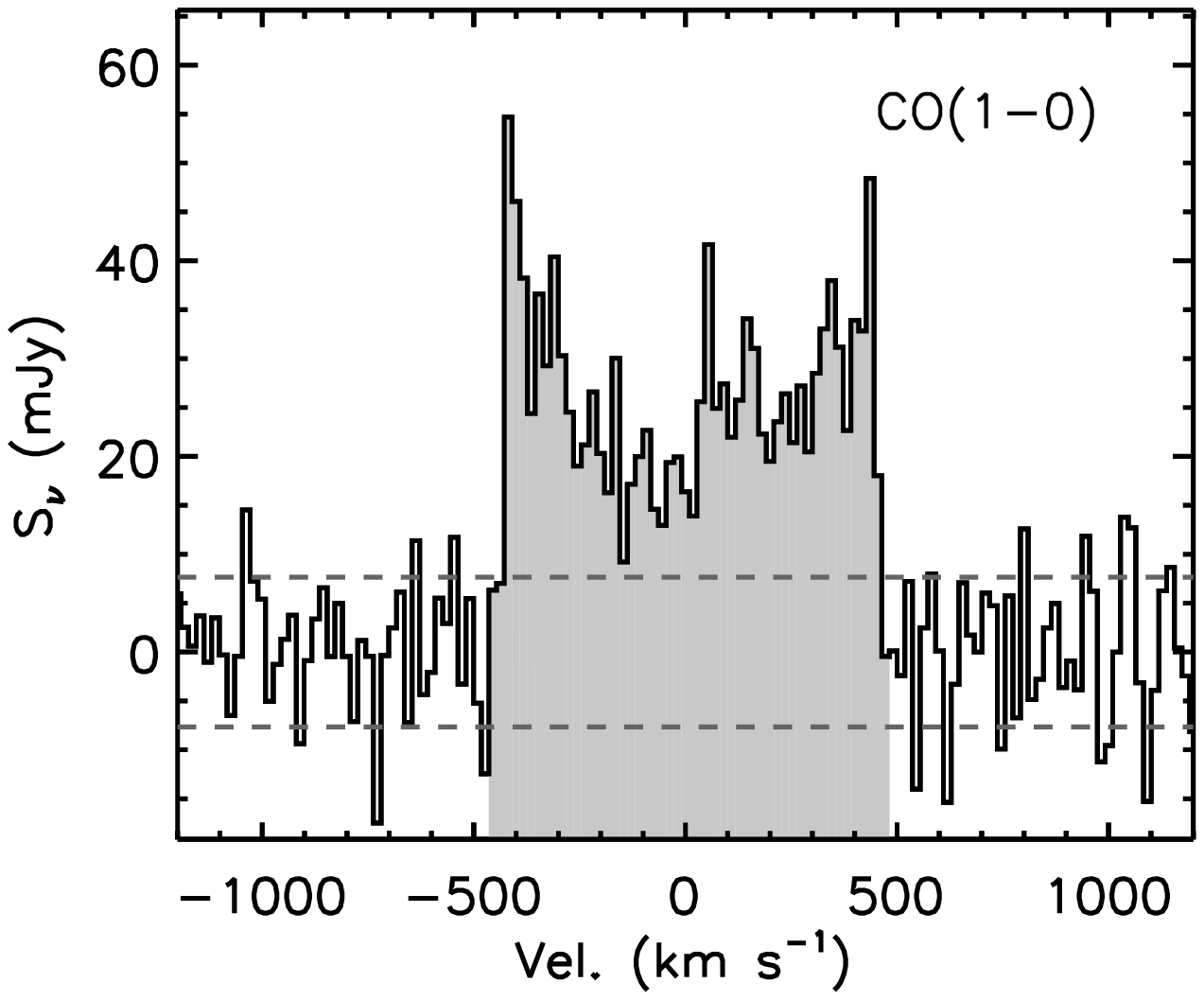} &
\includegraphics[width=5cm,angle=0,clip,trim=-0.2cm 0cm 0cm 0.0cm]{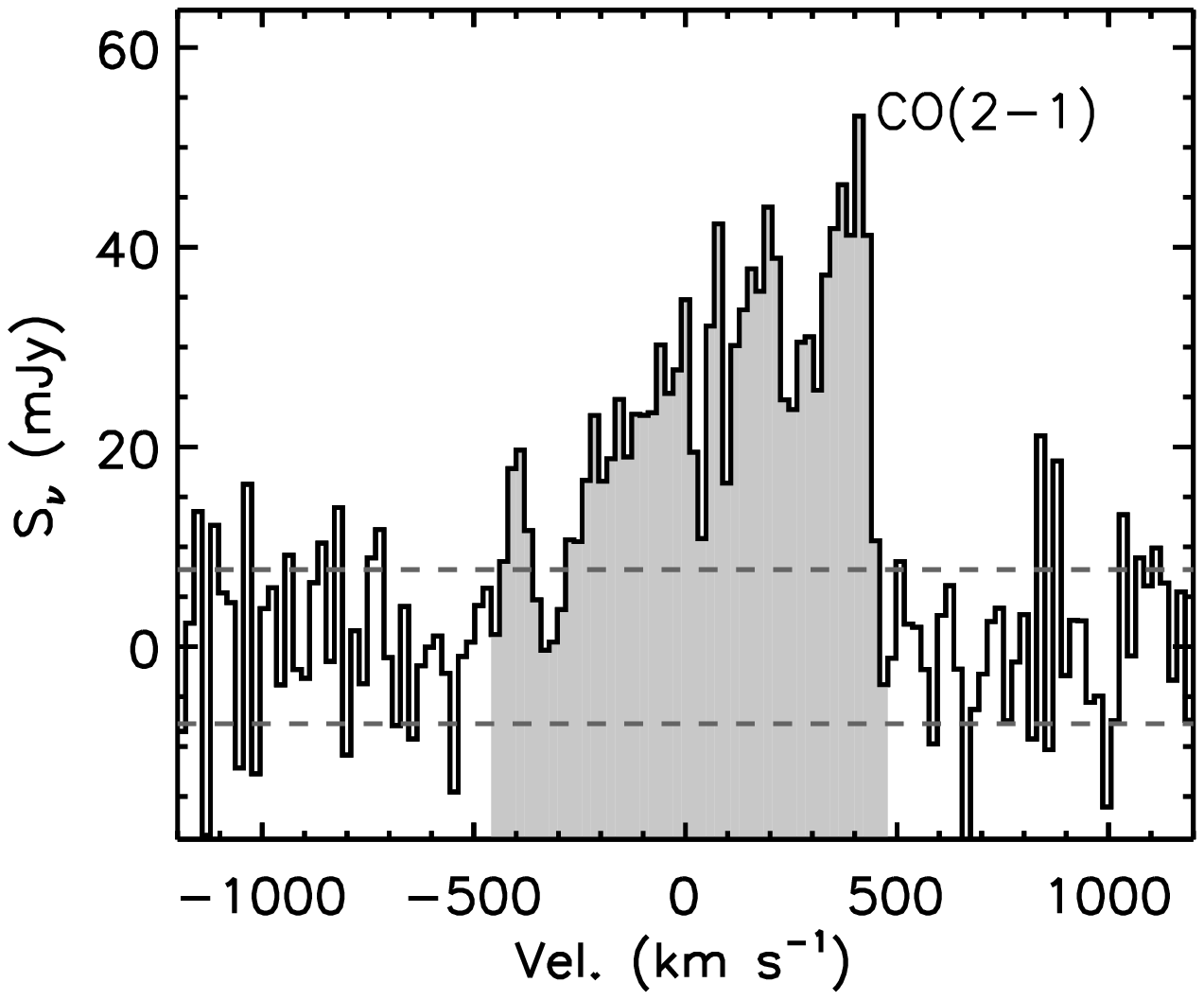} & 
\includegraphics[width=4.7cm,angle=0,clip,trim=0cm 0cm 0cm 0.0cm]{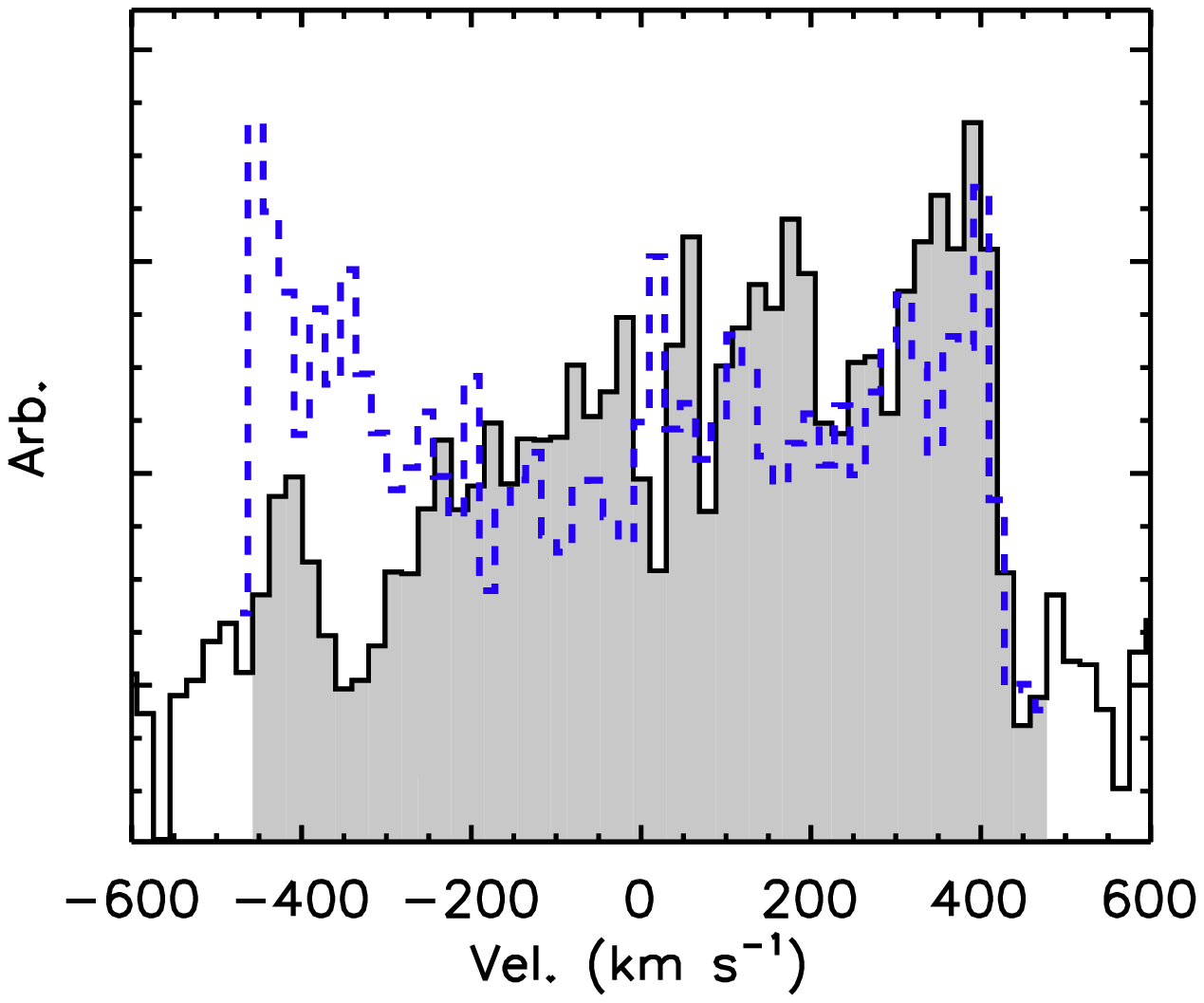} 
\end{array}$
 \end{center}
 \caption{CO(1-0) and CO(2-1) IRAM-30m spectra (left and middle column, respectively) for our detected sample galaxies. The velocity on the x-axis is plotted over $\pm$1200\kms\ with respect to the rest-frequency of that line. The grey shaded region on the spectra denotes the detected line, and the dashed lines show the $\pm$1$\sigma$ RMS level. The right column shows the line profiles over-plotted on one another (the CO(2-1) is a solid black line and CO(1-0) is a dashed blue line), scaled arbitrarily to the same peak brightness.}
 \label{codetsfig}
 \end{minipage}
 \end{figure*}

 \begin{table*}
\caption{Observational parameters and derived molecular gas masses for the sample ETGs}
\begin{tabular*}{\textwidth}{@{\extracolsep{\fill}}l r r r r r r r r r r r r r r}
\hline
Name & Peak$_{1\mhyphen0}$ & RMS$_{1\mhyphen0}$ & $\delta^{\rm chan}_{1\mhyphen0}$ & $W^{1\mhyphen0}_{20}$ & $\int S_{\nu}\, \delta V_{1\mhyphen0}$  & Peak$_{2\mhyphen1}$ &  RMS$_{2\mhyphen1}$ & $\delta^{\rm chan}_{2\mhyphen1}$ & $W^{2\mhyphen1}_{20}$ & $\int S_{\rm \nu}\, \delta V$$_{2\mhyphen1}$ & log$_{10}$(M$_{\rm H_2}$) \\ 
  & (mJy) & (mJy) & \multicolumn{2}{c}{(\kms)} & (Jy \kms) & (mJy) & (mJy) & \multicolumn{2}{c}{(\kms)} & (Jy \kms) & (M$_{\odot}$) \\
 (1) & (2) & (3) & (4) & (5) & (6) & (7) & (8) & (9) & (10) & (11) & (12)\\
\hline
IC0310 &       96.5 &       21.6 &        18. &       380. &       14.8$\pm$       1.8 &      334.6 &       46.9 &         5. &       340. &       43.9$\pm$       2.0 &       9.02$\pm$      0.05\\
NGC0665 &       94.8 &       20.4 &        16. &       450. &       23.0$\pm$       1.7 &      132.9 &       25.4 &        17. &       490. &       35.0$\pm$       2.3 &       9.18$\pm$      0.03\\
NGC0997 &       86.9 &       11.4 &        26. &       560. &       18.8$\pm$       1.4 &      227.5 &       34.8 &         6. &       540. &       42.6$\pm$       2.1 &       9.26$\pm$      0.03\\
NGC1132 & -- &        7.7 &        52. & -- & -- & -- &       13.2 &        52. & -- & -- & $<$      8.61\\
NGC1497 &       45.3 &       10.5 &        36. &       660. &       13.8$\pm$       1.7 &      122.5 &       23.2 &        19. &       660. &       35.7$\pm$       2.6 &       9.10$\pm$      0.05\\
NGC1684 &      117.8 &       14.3 &        21. &       820. &       33.6$\pm$       1.8 &      256.0 &       30.6 &         5. &       820. &       74.8$\pm$       2.0 &       9.20$\pm$      0.02\\
NGC2258 & -- &       13.1 &        52. & -- & -- & -- &       15.8 &        52. & -- & -- & $<$      8.81\\
NGC5208 &       54.7 &        7.7 &        18. &       950. &       23.7$\pm$       1.0 &       53.1 &        7.7 &        19. &       950. &       22.3$\pm$       1.1 &       9.49$\pm$      0.02\\
NGC5252 & -- &       10.6 &        52. & -- & -- & -- &       24.5 &        52. & -- & -- & $<$      8.86\\
NGC6482 & -- &        5.7 &        52. & -- & -- & -- &        6.8 &        52. & -- & -- & $<$      8.62\\
NGC7556 & -- &        4.3 &        52. & -- & -- & -- &        7.1 &        52. & -- & -- & $<$      8.56\\
\hline
NGC0708 & -- & -- &  -- &      560.& -- & -- &  -- &  -- & -- & -- &       8.83$\pm$      0.07\\
NGC1167 & -- & -- &  -- &      466.& -- & -- &  -- &  -- & -- & -- &       8.52$\pm$      0.12\\
NGC2320 & -- & -- &  -- &      830.& -- & -- &  -- &  -- & -- & -- &       8.41$\pm$      0.09\\
NGC7052 & -- & -- &  -- &      900.& -- & -- &  -- &  -- & -- & -- &       9.63$\pm$      0.06\\
         \hline
\end{tabular*}
\parbox[t]{1\textwidth}{ \textit{Notes:}  Column 1 lists the name of each source. Column 2 lists the peak CO(1-0) line flux for each object, and Column 3 the RMS noise reached in the observation with a channel size as shown in Column 4. Column 5 shows the CO(1-0) linewidth at 20\% of the peak intensity, and Column 6 the integrated intensity of the CO(1-0) line. Columns 7-11 show the same properties as Columns 2-6, but derived from the CO(2-1) line. All these are in units of mJy (converting these to an integrated line intensity in main beam temperature can be accomplished by dividing by a factor of 4.6 at 3mm and 5.95 at 1mm). Dashes indicate sources where the lines were not detected. Column 12 shows the logarithm of the H$_2$ mass for each source, derived using Equation \protect \ref{co2h2}. For the non detected sources Column 12 lists the 3$\sigma$ upper limit to the H$_2$ mass derived assuming a 250 \kms\ velocity width for the line. Below the divider we list H$_2$ masses (converted using our assumed X$_{\rm CO}$) and $W^{1\mhyphen0}_{20}$ values for MASSIVE galaxies detected elsewhere in the literature. NGC0708 and NGC7052 were detected by \cite{2010A&A...518A...9O}, NGC1167 by \cite{2015A&A...573A.111O} and NGC2320 by \cite{2005ApJ...634..258Y}.  }
\label{obstable}
\end{table*}

The IRAM 30-m telescope at Pico Veleta, Spain, was used between the 6th January and the 23rd March 2015 to observe CO emission in our sample galaxies (proposal 194-14, PI Greene/Davis), as part of the hetrodyne pool.   
 We aimed to simultaneously detect CO(1-0) and CO(2-1) (at rest-frequencies of 115.27 and 230.54 GHz respectively) in the 3mm and 1mm atmospheric windows.
The beam full width at half-maximum (FWHM) of the IRAM-30m at the frequency of these lines is 21\farc3 and 10\farc7, respectively. 
These beamsizes correspond to physical scales of between 5.6 and 9.8 kpc at the frequency of CO(1-0), and between 2.8 and 4.9 kpc at the frequency of CO(2-1), given the varying distance to these sources. 
 The Eight MIxer Receiver (EMIR) was used for observations in the wobbler switching mode, with reference positions offset by $\pm$100\arcsec\ in azimuth. The Fourier Transform Spectrograph (FTS) back-end gave an effective total bandwidth of $\approx$4 GHz per window, and a raw spectral resolution of 200 kHz ($\approx$0.6 \kms\ at 115GHz, $\approx$0.3 \kms\  at 230 GHz). The Wideband Line Multiple Autocorrelator (WILMA) back-end was used simultaneously with the FTS, and WILMA data were used at 230 GHz when possible to avoid known issues with platforming in the FTS. 

The system temperatures ranged between 152 and 235 K at 3~mm and between 230 and 270 K at 1~mm. The time on source ranged from 38 to 96 min, being weather-dependent, and was interactively adjusted by the pool observers to try and ensure detection of molecular emission.
 
The individual $\approx$6 minute scans were inspected, and baselined, using a zeroth-, first- or second-order polynomial, depending on the scan. Scans with poor quality baselines or other problems were discarded.
The good scans were averaged together, weighted by the inverse square of the system temperature. We consider emission lines where the integrated intensity has greater than a 3$\sigma$ significance (including the baseline uncertainty;  \citealt{2012MNRAS.421.1298C}) to be detected.  

Integrated intensities for each detected line in each galaxy were computed by fitting a Gaussian profile in the \texttt{\small CLASS} package of \texttt{\small GILDAS}\footnote{http://www.iram.fr/IRAMFR/GILDAS/ - accessed 14/01/13}, and by summing the spectrum over the velocity. Both methods produce consistent results, but as the observed profile shapes are non-Gaussian, we here present the values derived from the sum. 
We convert the spectra from the observed antenna temperature (T$_a^*$) into units of Janskies, utilising the point-source conversion as tabulated on the IRAM website\footnote{http://www.iram.es/IRAMES/mainWiki/EmirforAstronomers - accessed 23/04/2013}. The resulting spectra for the detected lines are shown in Figure \ref{codetsfig}, and the non-detections are shown in Figure \ref{conodetsfig} in Appendix A. Table \ref{obstable} lists the RMS noise levels in each spectrum, the velocity width summed over, and the line integrated intensities.

\section{Results}
\label{results}
\subsection{Line detections}

We detected line emission in six out of our eleven galaxies with a signal-to-noise of $>$3 (see Figure \ref{codetsfig}).  Table \ref{obstable} lists the properties of the detected lines. In each object we detect both the CO(1-0) and CO(2-1) lines, and these are both consistent with having the same systemic velocity as the galaxy itself.

The velocity width of a CO line in a galaxy depends both on the rotation curve of the galaxy, and the relative extent of the emitting components. 
The velocity linewidths found for our sources are very high (up to 950 \kms in NGC5208). We discuss this further in Section \ref{tfrsec} in the context of the CO Tully-Fisher relation (TFR; e.g. \citealt{2011MNRAS.414..968D}).

Real velocity width differences between different CO transitions may indicate that the emitting gas is concentrated differently or that the smaller beam at higher frequencies is missing some gas.  The velocity widths of the detected lines in our target ETGs are consistent within our errors in all cases, however the profile shapes are observed to differ.  The right column in  Figure \ref{codetsfig} shows this clearly, overplotting the CO(2-1) spectrum in blue over the CO(2-1) observation. In NGC5208 (and to a lesser extent in NGC1497) the CO(2-1) spectrum is highly asymmetric, while the CO(1-0) is not. While it is not possible to rule out differing intrinsic gas distributions, a far more likely explanation is that a small telescope pointing error causes gas from one side of the object to be missed in the smaller CO(2-1) beam, but to be included in the CO(1-0) observations. This allows us to roughly estimate that the size of the gas reservoir should be $>$11\arcsec, which corresponds to a physical size of $>$4.9 kpc in NGC5208.

In principle, the ratio of the CO(1--0) and CO(2--1) lines can be used to estimate the excitation temperature of the molecular gas, but the significantly smaller beam size of the IRAM-30m at the higher frequency of CO(2--1) acts as a systematic error in such a measurement. Thus the line ratio we observe is affected by both the excitation temperature and the spatial distribution of the gas.
 If the observed CO emission were to fill the telescope beam at both frequencies (and the CO is not sub-thermally excited) we would expect a line ratio of one (measured in main beam temperature units). However, if the CO emission is compact compared to the beam then the measured intensity in the CO(2--1) line should be larger by up to a factor of 4 (as the beam at such frequencies covers a 4 times smaller area). 
  In our objects the CO(2--1)/CO(1--0) main beam temperature ratio ranges from 1.2 to 2.29. NGC5208 formally has a lower ratio (0.78), but as argued above we expect to have missed significant CO(2-1) flux in this object due to a slight pointing offset. In the remaining cases this ratio is less than 2.5, suggesting the CO is reasonably extended, but still smaller than our 22\arcsec\ CO(1-0) beam.  This is what one would expect, based on the typical gas extent found in \atlas\ objects \citep{2013MNRAS.429..534D}.  The measured line ratios thus give us confidence that we are not missing significant amounts of molecular emission from larger radii in these objects, and thus the molecular masses derived below are robust. It should be noted however that sub-thermal excitation of CO lines would decrease the intensity of the higher frequency transition \citep[e.g.][]{1992A&A...264..433B}, so a firm statement about the extent of the gas is not possible without resolved data.   

\subsection{Gas masses}
\label{gasmasses}

\subsubsection{H$_2$}
We estimate H$_2$ gas masses for our CO detections in the standard manner, using the following equation

\begin{equation}
M_{H_2} = 2m_H \frac{\lambda^2}{2 k_b} X_{\rm CO} D_L^2 \int{S_v \delta V},
\end{equation}
\noindent where $m_H$ is the mass of a hydrogen atom, $\lambda$ is the wavelength, $k_b$ is Boltzmann's constant, $D_L$ is the luminosity distance, $\int{S_v \delta V}$ is the integrated CO flux density and  X$_{\rm CO}$ is your CO-to-H$_2$ conversion factor of choice in units of K \kms. 
For CO(1-0) this formula reduces to

\begin{equation}
\left(\frac{M_{H_2}}{\mathrm{M\odot}}\right) = 1.18\times10^8 \left(\frac{X_{\rm CO}}{3\times10^{20}\,\mathrm{K\, km\,s^{-1}}}\right) \left(\frac{D_L}{100\,\mathrm{Mpc}}\right)^2 \left(\frac{\int{S_v \delta V}}{\mathrm{Jy\,km\,s^{-1}}}\right).
\label{co2h2}
\end{equation}

Here we use a Galactic X$_{\rm CO}$ factor of 3$\times$10$^{20}$ cm$^{-2}$ (K \kms)$^{-1}$ \citep{Dickman:1986jz}. As these ETGs are massive, and mass correlates positively with metallicity, such a value seems reasonable. It is possible, however, that the gas in these galaxies has been accreted from a low-metallicity source. In such a case we would be underestimating the total gas mass. For non-detections we set upper limits on their H$_2$ mass using the 3$\sigma$ RMS on the CO(1-0) spectrum with 50 \kms\ channels, assuming a total velocity width of 250 \kms, to allow direct comparison to \citealt{2011MNRAS.414..940Y}). The detected MASSIVE objects all have larger velocity widths than this, and if we assumed a velocity width of 500 \kms\ our upper limits would increase by a factor of two.

\begin{figure}
\begin{center}
\includegraphics[height=8cm,angle=0,clip,trim=0cm 0cm 0cm 0.0cm]{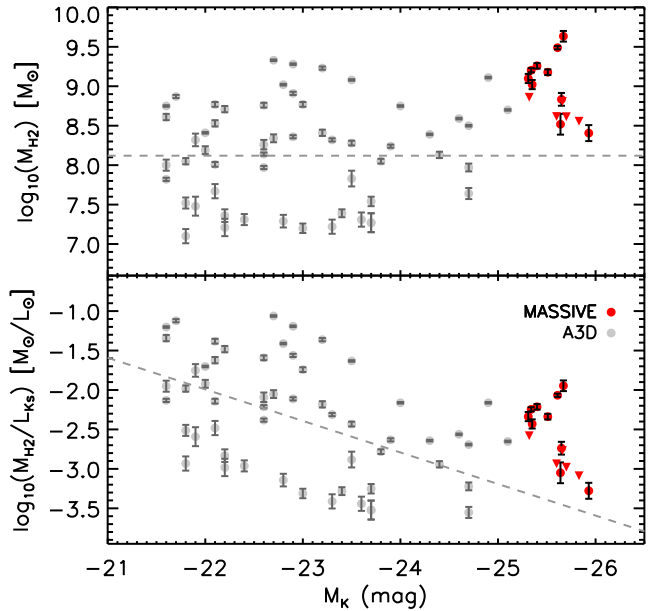}
 \caption{\textit{Top panel:} The base-10 logarithm of the molecular gas masses for the MASSIVE and \atlas\ ETGs, plotted as a function of $Ks$-band magnitude. \textit{Bottom panel:} The base-10 logarithm of the H$_2$-mass to $Ks$-band luminosity ratio (an observational proxy for gas fraction) plotted against the $Ks$-band magnitude. In both panels the MASSIVE objects are shown as red points, while \atlas\ galaxies are plotted in grey. The upper limits from this paper are shown as red triangles. The completeness limit of the \atlas\ survey is shown as a grey dashed line in both figures.}
 \label{h2_mass}
 \end{center}
 \end{figure}

We find that the galaxies newly observed in this work have H$_2$ masses between 1$\times$10$^9$ and 4.3$\times$10$^9$ \msun, while our 3$\sigma$ upper limits extend down to 3$\times$10$^8$ \msun. The value derived for each object is listed in Table \ref{obstable}. These H$_2$ masses are high, similar to the ISM mass found in spiral galaxies like the Milky Way. The top panel of Figure \ref{h2_mass} shows a comparison between the objects studied in this paper and the molecular-rich ETGs from the \atlas\ sample \citep{2011MNRAS.414..940Y}.

Our sample galaxies are on average $\approx$4 times as distant as the \atlas\ objects, and hence we are only sensitive to relatively high ISM masses (the upper limits also shown in Figure \ref{h2_mass} illustrate our detection limits, {and can be compared to the grey dashed line which is the completeness limit for \atlas}).
 In the (admittedly small) range of gas masses where the two samples overlap, there is slight evidence that the ISM masses present in these MASSIVE ETGs could be higher than those found in \atlas. For instance, the median gas mass {(for objects with $>$7$\times10^8$ \msun\ of gas, where both samples are complete)} of detected \atlas\ objects is  1$\times10^9$ \msun, while the median for MASSIVE is two times higher. Two of our objects also have higher ISM masses than any of the objects in \atlas. However, our sample is small and biased towards high gas masses, and in this overlapping region a Mann-Whitney U test is unable to reject at more than $1\sigma$ the hypothesis that the MASSIVE and \atlas\ objects are drawn from the same underlying distribution of H$_2$ masses. 

If the amount of H$_2$ in ETGs does not depend strongly on stellar mass, then one would expect a weak negative correlation to be present when one plots the molecular gas fraction (here simply M$_2$/M$_*$) versus stellar mass (as the typical amount of H$_2$ does not change, but the stellar mass increases, making the typical gas fraction smaller). 
Such a weak trend is observed in the bottom panel of Figure \ref{h2_mass}, where we plot the H$_2$-mass to $Ks$-band luminosity ratio (an observational proxy for gas fraction) against the $Ks$-band luminosity (a proxy for stellar mass). Our objects are again at the upper envelope of the correlation, which may be physical, or caused by our biased sample selection. 

\subsubsection{\hi}
Three of the detected galaxies have existing \hi\ mass measurements from the compilation of \cite{2005ApJS..160..149S}, and one has been interferometrically mapped by \cite{2010A&A...523A..75S}. We re-calculated the inferred \hi\ mass using our galaxy distances from Table \ref{proptable}, and find NGC0665 has M$_{\rm \hi}$ = 1.0 $\pm$ 0.3 $\times$10$^{9}$ M$_{\odot}$, NGC1167 has M$_{\rm \hi}$ = 1.5 $\pm$ 0.1 $\times$10$^{10}$ M$_{\odot}$, NGC1497 has M$_{\rm \hi}$ = 4.7 $\pm$ 1.5 $\times$10$^{8}$ M$_{\odot}$ and NGC5208 has M$_{\rm \hi}$ = 2.7 $\pm$ 0.9 $\times$10$^{9}$ M$_{\odot}$. Using the molecular gas masses derived in the previous Section, we find that these objects have H$_2$-to-\hi\ mass ratios of 1.4, 0.02, 2.6 and 1.1 respectively. With the exception of NGC1167 (which has a very large, low density \hi\ disc), this continues the trend seen in the \atlas\ survey that massive ETGs are often molecular gas dominated \citep{2012MNRAS.422.1835S}, possibly because of their deep potential wells and higher midplane hydrostatic-pressure \citep[e.g.][]{2002ApJ...569..157W}. 

\subsection{Star formation}

As discussed in the introduction, ETGs have relatively low star formation efficiencies (star-formation rate per unit gas mass) when compared to spiral and starburst galaxies \citep[e.g.][]{2011MNRAS.415...61S,2014MNRAS.444.3427D}. Some authors \citep[e.g.][]{2002PASJ...54..541K,2009ApJ...707..250M,2013MNRAS.432.1914M,2014MNRAS.444.3427D} have suggested this suppression is dynamical in nature, and is related to the depth of the potential well.
Our objects, by construction, are the most massive ETGs that exist locally, and thus it is interesting to see if such a suppression is also visible here.

In order to estimate the star formation rate (SFR) in these objects, we make use of \textit{WISE} \citep{2010AJ....140.1868W} 22 \mum\ data, as described in Section \ref{sfr_describe}.  
The derived star formation rates range from 0.07 M$_{\odot}$ yr$^{-1}$ in NGC2258 to 2.8 M$_{\odot}$ yr$^{-1}$ in NGC 5252. The star formation rate derived for NGC5252 is several times higher than those from any other source. We note that NGC 5252 has a strong Seyfert 2 AGN, and thus find it likely that emission from a hot torus is artificially boosting the derived SFR in this source. Discounting this galaxy the derived SFRs in detected galaxies are entirely consistent with those derived for other populations of ETGs studied in the literature \citep{2007MNRAS.377.1795C,2011MNRAS.415...61S,2013MNRAS.432.1914M,2014MNRAS.444.3427D,2015MNRAS.449.3503D}, with a median SFR of 0.28 M$_{\odot}$ yr$^{-1}$. We note that any AGN contribution in the other objects would make these SFRs upper limits (and hence our conclusions below remain robust).

\begin{figure}
\begin{center}
\includegraphics[width=0.48\textwidth,angle=0,clip,trim=0cm 0cm 0cm 0.0cm]{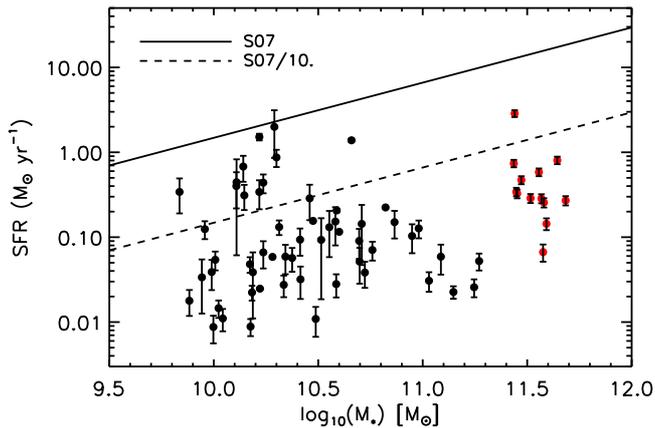}
 \caption{Star formation rate versus stellar mass in our 10 detected  MASSIVE ETGs (red points) and in the \atlas\ ETGs (black points). The star formation rate `main-sequence' of \protect \cite{2007ApJS..173..267S} is shown as a black line, while the dashed line indicates where objects lie an order of magnitude below that relation.}
 \label{ssfrplot}
 \end{center}
 \end{figure}

We can estimate a stellar mass for these galaxies using the $K$-band magnitude and assuming a $K$-band mass-to-light ratio of one (a-priori reasonable for an old stellar population; e.g. \citealt{2001ApJ...550..212B}). Using this mass, we find that these objects have specific star formation rates (sSFR=SFR/M$_{*}$) of between $1\times10^{-11}$ and $2\times10^{-13}$ yr$^{-1}$. In Figure \ref{ssfrplot} one can see that, as expected, these objects all fall below the low redshift ``star-forming galaxy main sequence'' \citep[e.g.][]{2004MNRAS.351.1151B}. They lie at similar sSFR as the \atlas\ objects, in the quenched galaxy regime (consistent with their red optical colours), despite some having star formation rates as high as the Milky Way. Our MASSIVE objects appear to have significantly higher SFR than the most massive ETGs in the \atlas\ survey (e.g. those with stellar masses $>$10$^{11}$ \msun). This seems likely to arise because of our biased selection towards high SFR objects, but this will be confirmed in future works. 

We are unable to place our galaxies on the Kennicutt-Schmidt relation \citep[e.g.][]{1998ApJ...498..541K}, as we have no estimate of the size of the star-forming region in these objects. We can however estimate the star-formation efficiency (SFE). In what follows we define the SFE as star-formation rate per unit molecular gas mass (SFR/M$_{\rm H2}$) rather than total gas, because molecular gas is the phase which is most closely linked to star formation \citep[e.g.][]{2008AJ....136.2846B}. Including \hi\ (where available) in our SFEs would make them smaller (or equivalently, make depletion times longer). We note that the preselection of our pilot sample objects, that predisposes them to have large SFRs, should not effect study of the SFE as the normalisation by the molecular gas mass should minimise any bias.

\begin{figure}
\begin{center}
\includegraphics[width=0.45\textwidth,angle=0,clip,trim=0cm 0cm 0cm 0.0cm]{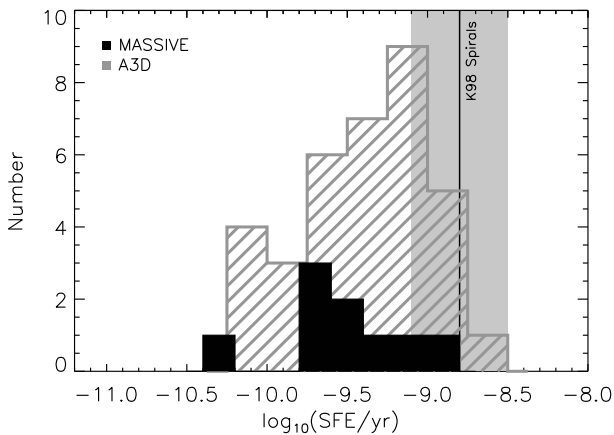}
 \caption{Histogram of the star formation efficiency (SFE=SFR/M$_{\rm H2}$) in our 10 detected  MASSIVE ETGs (black histogram) and \atlas\ ETGs (grey shaded histogram). The average SFE of the spiral galaxies from \protect \cite{1998ApJ...498..541K} is also shown as a black vertical line, with a grey bar showing the standard deviation of their SFEs around the mean. }
 \label{sfeplot}
 \end{center}
 \end{figure}

Figure \ref{sfeplot} shows the SFE in our CO detected galaxies (black histogram), and the molecule-rich \atlas\ ETGs (grey histogram). Also shown (as a black line/grey bar) are the spiral galaxies of \cite{1998ApJ...498..541K}, which have been found to have average SFEs of $\approx$1.5$\times10^{-9}$ yr$^{-1}$ (see also e.g. \citealt{2011ApJ...730L..13B}). The typical SFE of our galaxies is substantially lower than found in spiral objects, but overlaps with the distribution found in \atlas. A Mann-Whitney U test is unable to reject the hypothesis that the MASSIVE and \atlas\ objects are drawn from the same underlying distribution. 
We note that if one were to neglect the correction for circumstellar emission, the star formation rates of our sources would increase, but the average star formation efficiency of these objects would still be lower than found in spiral galaxies.
Overall we thus find evidence of star-formation suppression in these MASSIVE galaxies. If star formation were to continue on in a steady state at the same level as today, the gas reservoir in our objects would take a median time of $\approx$4.6 Gyr to be depleted.

\subsection{The CO Tully-Fisher relation}
\label{tfrsec}

\begin{figure*}
\begin{center}
\includegraphics[height=8cm,angle=0,clip,trim=0cm 0cm 0cm 0.0cm]{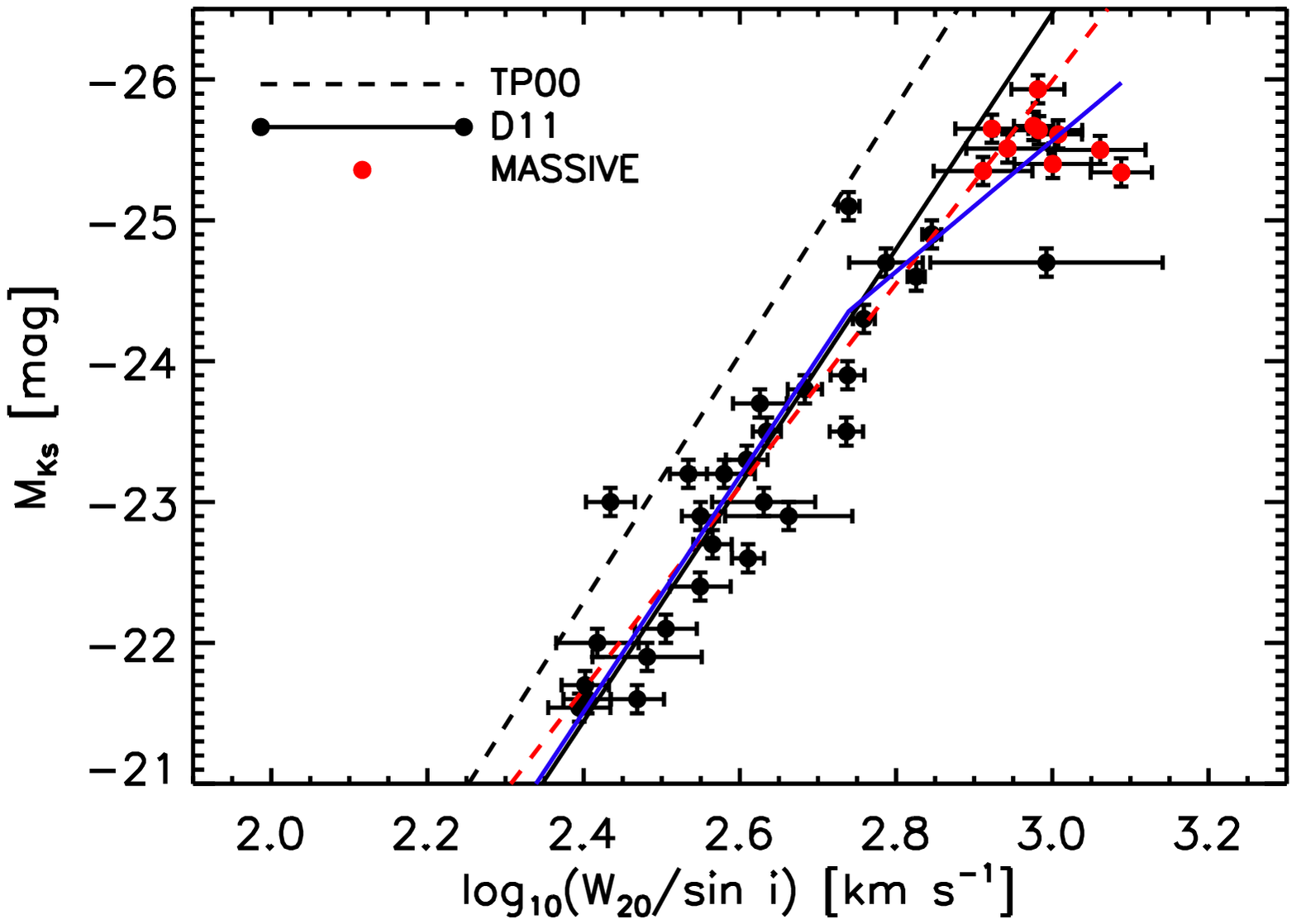}
 \caption{The CO Tully-Fisher relation for the 10 detected MASSIVE, and the \atlas ETGs. The logarithm of W$_{20}$, the measured velocity width at 20\% of the peak height for the CO line (corrected for the galaxy inclination), is plotted as a function of absolute $K$-band magnitude. MASSIVE objects are shown in red, while \atlas\ ETGs are shown in black. The dashed black line shows the spiral galaxy TFR from \protect \cite{2000ApJ...533..744T}, while the solid black line is the best fit for ETGs from \protect \cite{2011MNRAS.414..968D}. Our best fit single power law is shown as a red dashed line, while the best fit broken power law is shown in blue.}
 \label{tfr}
 \end{center}
 \end{figure*}

The Tully-Fisher relation (TFR; \citealt{1977A&A....54..661T}) is a key scaling relation for spiral
galaxies. The underlying cause of this relation between
the luminosity and rotational velocity is usually interpreted as the
product of a relatively constant total (luminous plus dark) mass-to-light
ratio (M/L) in the local spiral galaxy population (\citealt{1993ApJ...419..469G,1995MNRAS.273L..35Z}), and hence a strong coupling between
the dark and luminous masses. Studying the slope and zero-point
of the TFR is thus also a powerful probe of the M/L evolution of
galaxies (e.g. \citealt{1989A&A...211..259P,1995ApJ...438...72S,2001ApJ...550..212B}).

The TFR of ETGs is also important, as the TFR is likely to encode clues
about their assembly and evolution. For example, many authors
have suggested that S0 galaxies have avoided violent interactions
and are the faded descendants of high-redshift spirals (e.g. \citealt{1980ApJ...236..351D,1997ApJ...490..577D}). In this scenario, galaxies become dimmer
whilst keeping the same dynamical mass, leading to an offset
TFR. Recent analyses of ETGs have
suggested that both lenticular and elliptical galaxies do indeed have a measurable offset from the
spiral TFR, of around 0.5 -- 1.5 mag at K band \citep{1999AJ....117.2666N,2001MNRAS.322..702M,2001AJ....121.1936G,2006MNRAS.373.1125B,2007ApJ...659.1172D,2010MNRAS.409.1330W,2013MNRAS.432.1010C,2013MNRAS.433.2667R,denHeijer2015}.
Fading alone is not expected to fully explain this offset, however, suggesting size evolution of ETGs is also needed \citep{2010MNRAS.409.1330W}.

The possibility of using CO linewidths to investigate the TFR was
first explored by \cite{1992ApJ...393..530D} and this method has since
been used by various authors to investigate the CO-TFR of spiral/irregular galaxies (e.g. \citealt{1997A&A...323...14S,1994A&A...283...21S,1997A&A...323...14S,1997A&A...326..915T,1999A&A...351..467T,1998AJ....116.2672L,2001PASJ...53..701T}), quasars
\citep{2007ApJ...669..821H}, and ETGs \citep{2011MNRAS.414..968D}.

The TFR of these very massive galaxies is also interesting in its own right. 
Some authors have suggested that there is a change in the slope
of the TFR for high-mass galaxies, brighter than an absolute
K-band magnitude of $\approx$-23.75 mag (\citealt{1993ApJ...418..626P,2001ApJ...563..694V,2001AJ....121.1936G,2007MNRAS.381.1463N,2007ApJ...659.1172D}). It has been suggested that this
break occurs because many massive galaxies have centrally peaked rotation curves, which decline towards larger radii, whereas low-mass galaxies have relatively
flat circular velocity curves. 

As molecular gas has a typical radial extent of 1-2 kpc in ETGs \citep{2013MNRAS.429..534D}, it primarily traces the inner parts of the potential (where dark matter is negligible) and hence is sensitive to the inner mass profile shape. 
\cite{2011MNRAS.414..968D} found some tentative evidence of a break at high mass in the CO TFR of \atlas\ ETGs, but these objects seemed to show flat rotation curves (at least at small radii). Here we use the objects from this paper to extend the work of \cite{2011MNRAS.414..968D} to higher masses, in an attempt to confirm or refute the presence of a break in the TFR.

In Figure \ref{tfr} we plot the absolute $Ks$-band magnitude of our objects, versus the measured velocity width at 20\% of the peak height (W$_{20}$; essentially identical to the total velocity width in these sources, because of their sharp edged double horn line profiles) for the CO(1-0) line (red points). 
The velocity widths have been corrected for the estimated galaxy inclination, as derived in Section \ref{inc_est}. If the molecular gas is misaligned from the stellar body then these estimates of the inclination will be incorrect. Using the data we have available we suspect this may be the case in NGC1497 (as the dust lane visible in \textit{HST} imaging is orientated along the optical minor axis of the galaxy). Unless the potential of NGC1497 is symmetric, the gas will not rotate with the same circular speed, and would not be expected to follow the TFR. We do not include this object in the rest of the analysis in this Section, but including it would not change our main results.

We also show in Figure \ref{tfr} the lower-mass ETG objects from the `hybrid' sample of \citet[where the best inclination method available has been used to de-project the line-widths of every object]{2011MNRAS.414..968D} as black points. These authors used an identical method to derive the TFR parameters. As a black dashed line we show the spiral galaxy TFR from \cite{2000ApJ...533..744T} (derived using \hi\ linewidths). Clearly the ETGs are offset from this relation, by $\approx$1 magnitude. The best fit to the black points from \cite{2011MNRAS.414..968D} is shown as a black solid line. The MASSIVE galaxies lie systematically to the right (towards dimmer magnitudes/faster rotation) of the best fit for the \cite{2011MNRAS.414..968D} sample. 

It is hard to say if the addition of these objects at higher mass confirm that the TFR is a broken power law (as suggested by other authors), or if the ETG TFR simply has a different slope from that found for spiral galaxies. We here fit two models to the data. The first is a single power law of the standard form:

\begin{equation}
\label{fiteq}
{\rm M}_{K} =  m\,\left[{\rm log}_{10}\left(\frac{W_{20} \sin^{-1} i}{\rm km\ s^{-1}}\right) - 2.6\right]+c,
\end{equation}
where m is the slope and c is the zero-point of the relation. 

The second model is a broken power law, that exhibits the following behaviour:

\begin{eqnarray}
\label{fiteq2}
{\rm M}_{K}  =
\left\{
	\begin{array}{ll}
		-8.38\,x+c  & \mbox{if } W_{20} \leq W_{\rm break} \\
		m_2\,x+f(m,c,W_{\rm break})  & \mbox{if } W_{20} > W_{\rm break}
	\end{array}
\right.
\\
\mathrm{where}\,\, x = {\rm log}_{10}\left(\frac{W_{20}\sin^{-1} i}{\rm km\ s^{-1}}\right) - 2.6. \nonumber
\end{eqnarray}
At less than some critical value (W$_{\rm break}$) this function has a slope fixed to that found by \cite{2011MNRAS.414..968D} with a free intercept. Above this value a second power law with a free gradient is fitted. The two power laws are constrained to meet at the threshold value, meaning the intercept of the second power law is determined uniquely by W$_{\rm break}$, m$_2$ and c. We fit these two models to the data using a Markov Chain Monte Carlo approach, that allows us to obtain the full Bayesian posterior probability distribution for the fitted parameters (which were all given flat priors). The marginalised best fit parameters can be found in Table \ref{tfrfits}, along with the reduced chi-square of the best fit. Note that both reduced chi-square values are substantially greater than one, suggesting (as found by other authors) that the TFR has an intrinsic scatter of 0.3-0.4 dex.

A broken power law model clearly fits our combined dataset better than a single, shallow TFR, with a reduced chi-square (chi-square per degree of freedom) that is somewhat lower. However, as discussed in detail in \cite{2010arXiv1012.3754A} this does not on its own mean that a broken TFR is the best physical model. 
Below we present further evidence that this offset may be physical, and speculate about its cause.

\subsection{Potential causes of a broken TFR}

At its heart, the Tully Fisher relation connects two different tracers of the galaxy potential, one measured directly using dynamics, and the other indirectly via stellar luminosity. As CO kinematics allow us to only probe the inner regions of ETGs, where dark matter is expected to be negligible (see e.g. \citealt{2013MNRAS.432.1709C}), in order to cause a break in the TFR one must change the baryonic component of galaxies. One can:

\begin{enumerate}
\item have a systematic bias in the observational quantities for high mass objects
\item change the mass-to-light ratio of the stellar population (M$_*$/L$_*$) in the MASSIVE objects, decreasing their luminosities, while keeping their total mass the same.
\item decrease M$_*$/L$_*$ in the MASSIVE objects, while increasing the total stellar mass of the galaxy
\item keep M$_*$/L$_*$ the same, but change the shape of the galaxy potential in MASSIVE objects. Placing more mass at the galaxy centre would increase W$_{20}$ while holding the integrated luminosity constant. 
\end{enumerate}

The first point, above, requires some systematic bias that affects the measurement of W$_{20}$ or M$_{K_s}$.  In order for W$_{20}$ to be be biased, it would require the CO to be disturbed, the inclination to be wrong, or that the CO systematically probes different parts of the potential in the more massive ETGs. The regular line profiles seen in Figure \ref{codetsfig} argue strongly against the gas reservoirs of these objects being systematically disturbed. 
A bias based on gas probing different parts of the potential would require the CO to systemically reach the peak of the rotation curve in our MASSIVE sample objects, but to fail to do so in lower mass objects.
However, interferometric evidence presented in  \cite{2011MNRAS.414..968D} show that the CO reaches beyond the peak of the rotation curve in the vast majority of low mass ETGs, and thus we consider this explanation unlikely. A systematic inclination error is possible, as more massive ETGs tend to be rounder, meaning we would underestimate their true inclination. However, in the objects where dust-lanes are visible (or resolved gas interferometry allows direct measures of the gas inclination) we still observe an offset.

A systematic bias in the $K$-band photometry is possible, and indeed various authors \citep[e.g.][]{2007ApJ...670..249L,2012PASA...29..174S} have argued that such a bias can exist for high-mass galaxies. For the MASSIVE sample in particular, tests using deeper $V$-band data revealed little bias in the selection of survey galaxies \citep{2014ApJ...795..158M}, and thus we do not expect problems in the photometry to be substantially affecting the TFR we derive. Deeper $K$-band photometry is, however, being collected as part of the survey, which will firmly establish this at a future date.

As we are able to argue against an observational problem causing the offset TFR for high-mass ETGs, we below consider how to distinguish if the M/L or the total mass is varying in our objects. In order to do this, we take advantage of another scaling relation of ETGs, the Faber-Jackson relation (FJR; \citealt{1976ApJ...204..668F}). The FJR connects the stellar velocity dispersion ($\sigma$; another dynamical tracer of the potential) with the luminosity of the galaxy. 

If the mass distribution within massive ETGs changes systematically (either globally, or with more mass being placed at small radii) this will increase the circular velocity (here traced using the CO emitting gas) and the velocity dispersion of the stars in a similar manner. In such a case the residuals around the TFR should correlate with stellar velocity dispersion. However if the mass-to-light ratio of the stellar population changes for some reason, without the distribution and mass of stars being affected, then the mass profile of the object would be unchanged and both W$_{20}$ and $\sigma$ will remain the same.

\begin{figure}
\begin{center}
\includegraphics[width=0.5\textwidth,angle=0,clip,trim=0cm 0cm 0cm 0.0cm]{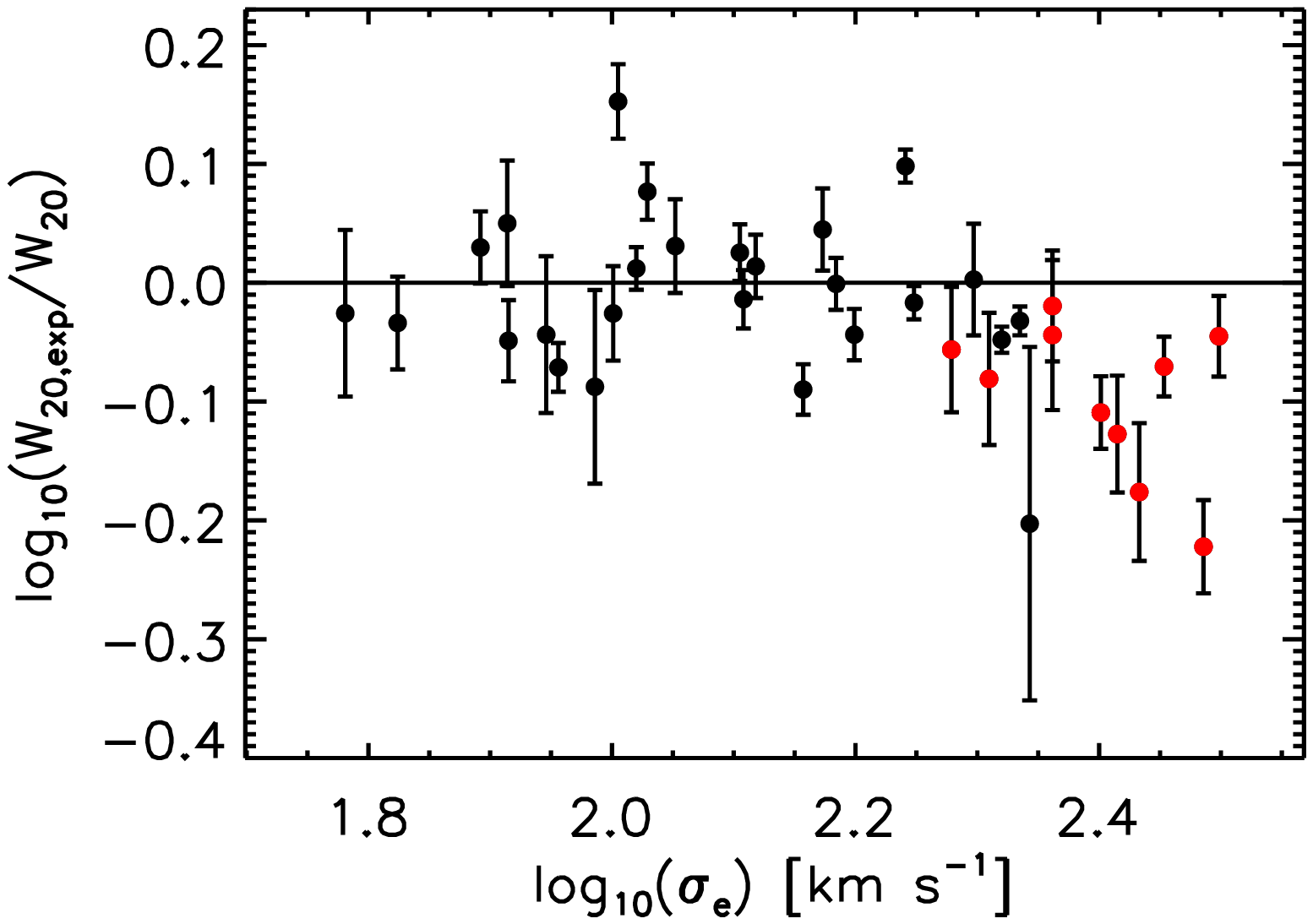}
 \caption{The residuals around the CO TFR of \protect \cite{2011MNRAS.414..968D} for the \atlas\ and MASSIVE galaxy sample, plotted as a function of the stellar velocity dispersion. MASSIVE objects are shown in red, while \atlas\ ETGs are shown in black. For \atlas\ the velocity dispersions are uniformly measured within an aperture of one effective radius (and are thus true measures of $\sigma_e$). In MASSIVE we take the values available in the literature, as tabulated in \protect \cite{2014ApJ...795..158M}.}
 \label{tfr_resids}
 \end{center}
 \end{figure}

In Figure \ref{tfr_resids} we show the velocity residuals around the best fit TFR of \cite{2011MNRAS.414..968D}, plotted as a function of stellar velocity dispersion. Above a velocity dispersion of $\approx$200 \kms\ the residuals around the TFR appear to correlate with $\sigma$. For \atlas\ the velocity dispersion is measured uniformly using IFS data over an aperture enclosing the effective radius of the galaxy. However, for MASSIVE the IFS data is not yet available, and we are forced to use in-homogenous literature values (as tabulated in \citealt{2014ApJ...795..158M}). 
If we were to use the homogenous central velocity dispersion values available for some of these objects from \cite{2015ApJS..218...10V}, however, this would not change our conclusions. {Given these uncertainties, and the low number of objects in this study this correlation is weak, with a Spearman's rank correlation coefficient of -0.36 (which deviates from the null-hypothesis by $\approx$1$\sigma$).}

If real, a correlation between an individual galaxy's velocity dispersion and its offset from the TFR would imply that the offset is physical in nature, and likely caused by a changing the amount of mass in the central regions of the most massive ETGs. It is not possible to determine from this simple data if such a change is global (and accompanied by a change in the stellar M/L in order to not change the total luminosity of the objects) or local (the effect of changing the mass profile in massive ETGs). In addition, we caution that other explanations are possible (e.g. the increased importance of triaxiality in massive ETGs).

With further observations however, one should be able to uniquely distinguish the cause of the offset seen in the TFR. For instance, resolved observations of the gas discs, and/or IFS observations of the stellar kinematics in these objects will reveal the circular velocity profiles of these MASSIVE galaxies, allowing a direct determination of the inclination of the gas discs, and study of M/L variations thorough stellar population analyses. 
If the offset TFR is caused by a changing mass profile then the offset should persist when one considers the total mass (baryonic) TFR, while if the cause was a mass-to-light ratio variation then the offset would be removed. We lack the stellar population information required to accurately convert our $K$-band photometry to total stellar mass (and have a lack of information on the \hi\ content of these galaxies), and hence postpone this analysis to a future work.

 \begin{table}
\caption{Best fitting TFR parameters}
\begin{tabular*}{0.48\textwidth}{@{\extracolsep{\fill}}l r r r r r}
\hline
Form & m & c & log$_{10}$(W$_{\rm break}$) & m$_2$  & $\chi^2_{\rm red}$\\ 
  &  & (mag) & (\kms) &  & \\
 (1) & (2) & (3) & (4) & (5) & (6)\\
\hline
\\
Single &  -7.21$_{-0.3}^{+0.17}$ & -23.11$_{-0.06}^{+0.03}$ & -- & -- & 3.95\\
\\
Double & -- & -23.19$_{-0.03}^{+0.09}$ & 2.74$_{-0.03}^{+0.07}$ & -4.7$_{-0.85}^{+0.74}$ & 2.85\\
\\
        \hline
\end{tabular*}
\parbox[t]{0.48\textwidth}{ \textit{Notes:}  The best parameters for a single power law, and double broken power law relation (defined in Equations \ref{fiteq} and \ref{fiteq2}) fitted to the data shown in Figure \ref{tfr}. The errors shown are the marginalised 1$\sigma$ uncertanties returned by the MCMC fitting routine. The reduced chi-square (the chi-square per degree of freedom) is shown in column 6.}
\label{tfrfits}
\end{table}

\section{Discussion \& Conclusions}
\label{discuss}
\label{conclude}

In this work we study CO(1-0) and CO(2-1) observations of 15 early-type galaxies in a previously unexplored mass range, drawn from the MASSIVE survey \citep{2014ApJ...795..158M}.
 These objects were selected because they showed signs of an ISM and/or star-formation, and thus our detection rate is high ($\approx$66\%; 10/15). The detected objects all have $>$2$\times$10$^8$ \msun\ of molecular gas present (assuming a Galactic conversion between CO and H$_2$ mass), and our upper limits in the undetected objects reach a factor of 3 deeper. 

Given the shallow nature of this pilot study, and the limited number of objects, it is hard to draw strong conclusions about the demographics of molecular gas in the most massive galaxies. 
In the objects where gas reservoirs exist, we find that the amount of molecular gas present is high, but are unable to fully reject the hypothesis that the amount of molecular gas present is independent of galaxy mass (as found for lower mass ETGs in previous works). 
A molecular gas mass distribution which is independent of galaxy mass has been interpreted as evidence that the supply of gas in ETGs dominated by mergers (and accretion of \hi\ stripped from nearby galaxies in interactions), rather than gas returned to ISM via stellar winds (stellar mass loss) or hot halo cooling (see e.g. \citealt{2011MNRAS.417..882D}). 
The tentative evidence of a higher gas mass in MASSIVE ETGs may reflect that stellar mass loss and cooling are starting to play a role in the most massive ETGs.
Further investigation of a more complete sample of massive ETGs is clearly needed, however, to firmly determine the fate of stellar mass loss and cooled material in massive galaxies.

\cite{2011MNRAS.414..940Y} found that molecular gas was usually only present in fast-rotating ETGs. Given that the fraction of slow rotators is significantly higher at high stellar mass \citep{2011MNRAS.414..888E}, the parent MASSIVE sample should contain a reasonable number of such objects. We currently lack the data to reliably determine the kinematic classification of these pilot sample galaxies individually. As the MASSIVE survey progresses, however, we will be able to tell if the result of \cite{2011MNRAS.414..940Y} holds, and all CO detected objects are fast rotators, or if the \atlas\ survey volume simply did not contain enough slow rotators.

We investigated the star formation properties of the target galaxies, using \textit{WISE} 22 \mum\ luminosities. After correction for the old stellar contribution at 22 \mum, we find that our objects have SFRs between 0.07 and 2.8 M$_{\odot}$ yr$^{-1}$,  with a median of 0.28 M$_{\odot}$ yr$^{-1}$, typical of other ETG samples. The sSFR of these objects is low, consistent with the red optical colours of the most massive ETGs. As highlighted by \cite{2014MNRAS.444.3408Y} it is easy to hide significant amounts of star formation in massive objects, whose colours will remain dominated by the old stellar population. 

These ETGs seem to have lower star-formation efficiencies than spiral galaxies, and the SFE's derived are consistent with being drawn from the same distribution as those previously found in other ETG samples. This suggests that the SFE is not simply a function of stellar mass, but that more local, internal processes ongoing in the galaxy are important to regulate star formation. These could be related to the stability of the gas in a steep potential \citep[e.g.][]{2002PASJ...54..541K,2009ApJ...707..250M}, related to local dynamical processes such as shear \citep[e.g.][]{2014MNRAS.444.3427D}, or more exotic in origin. Only with resolved data will we be able to tell what is suppressing star formation in these most massive ETGs.

Finally, we used our observed CO line profiles to investigate the high mass end of the Tully-Fisher relation. Previous studies had suggested that the TFR of ETGs is broken at an absolute K-band magnitude of $\approx$-23.75 mag. We showed that these MASSIVE ETGs do indeed lie offset from the best fit CO TFR fitted to lower mass ETGs. Our best fit break happens at an absolute K-band magnitude of $\approx$-24.4$^{+0.21}_{-0.54}$ mag, consistent within 2$\sigma$ with the break suggested in previous studies. 
In absolute terms it is difficult to tell if the TFR is broken, or simply shallower than previously thought.
A weak correlation observed between an individual galaxy's stellar velocity dispersion and its offset from the TFR, however, suggests that the offset is physical in nature, and likely caused by additional mass in the central parts of the most massive ETGs. 

This supports the suggestion of \cite{2007MNRAS.381.1463N} (revisited in \citealt{denHeijer2015}) that massive ETGs have steep rotation curves, which peak and then fall at larger radii. Thus integrated linewidths overestimate the true asymptotic halo velocity. 
The break in the TFR we observe here does not seem to support the suggestion that ETGs have a universal power-law slope total mass density profile (as suggested by e.g. \citealt{2015ApJ...804L..21C}), as in objects with universal density profiles the TFR offset would be constant with stellar mass.

As we find evidence for an increased offset in our MASSIVE ETGs, it is possible that high-mass ETGs have different formation histories that lead to increased central luminous mass concentrations (and a breakdown of the ``disk-halo conspiracy'';  \citealt{Kent:1987hy,Gavazzi:2007bl}). With further CO observations of massive ETGs, and resolved studies (with e.g. the NOrthern Extended Millimeter Array; NOEMA, or the Atacama Large Millimeter Array; ALMA) will we be able to determine what is causing the offset in these ETGs, if this persists in all objects of this class, and thus the implications for galaxy formation theories.

 \hspace{2.5cm}
 
 \vspace{0.5cm}
\noindent \textbf{Acknowledgments}

TAD acknowledges support from a Science and Technology Facilities Council Ernest Rutherford Fellowship, and thanks Freeke van de Voort for comments which improved the manuscript. The MASSIVE survey is supported in part by NSF AST-1411945 and AST- 1411642.
 This paper is based on pool observations carried out with the IRAM Thirty Meter Telescope, and we thank Claudia Marka, the pool supervisor for observational assistance. IRAM is supported by INSU/CNRS (France), MPG (Germany) and IGN (Spain). 

This publication makes use of data products from the Wide-field Infrared Survey Explorer, which is a joint project of the University of California, Los Angeles, and the Jet Propulsion Laboratory/California Institute of Technology, funded by the National Aeronautics and Space Administration. This research also made use of the NASA/IPAC Extragalactic Database (NED) which is operated by the Jet Propulsion Laboratory, California Institute of Technology, under contract with the National Aeronautics and Space Administration.

\bsp
\bibliographystyle{mnras}
\bibliography{bibMASSIVEpilot.bib}
\bibdata{bibMASSIVEpilot.bib}
\bibstyle{mnras}

\label{lastpage}

\clearpage
\appendix

    \begin{figure*}
    \section{CO non-detections}
\begin{minipage}[t]{1.0\textwidth}
\begin{center}
$\begin{array}{ccc}
\begin{turn}{90}\large \hspace{1.1cm} NGC1132\end{turn} &
\includegraphics[width=5cm,angle=0,clip,trim=0.0cm 1.3cm 0cm 0.0cm]{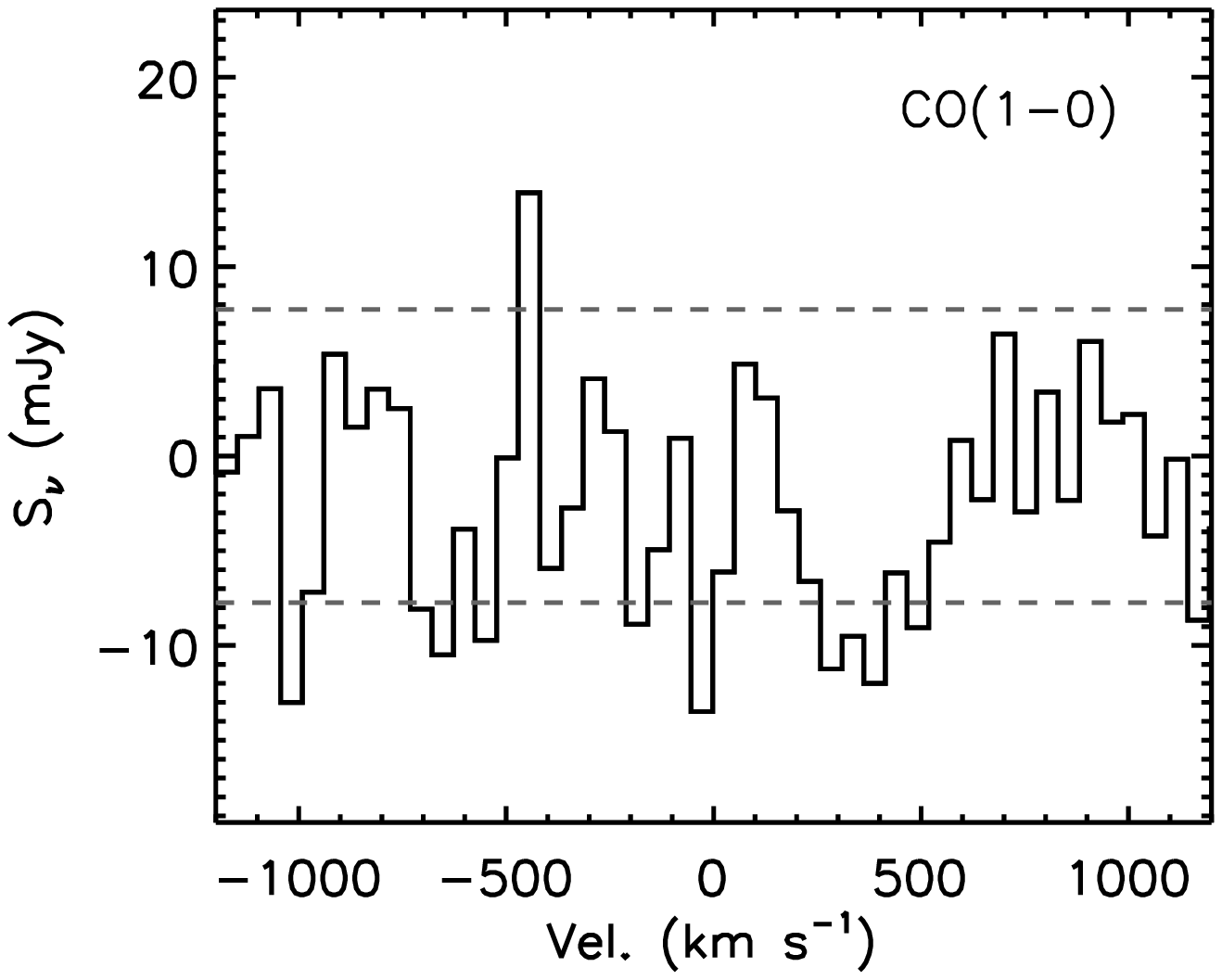} &
\includegraphics[width=5cm,angle=0,clip,trim=0.0cm 1.3cm 0cm 0.0cm]{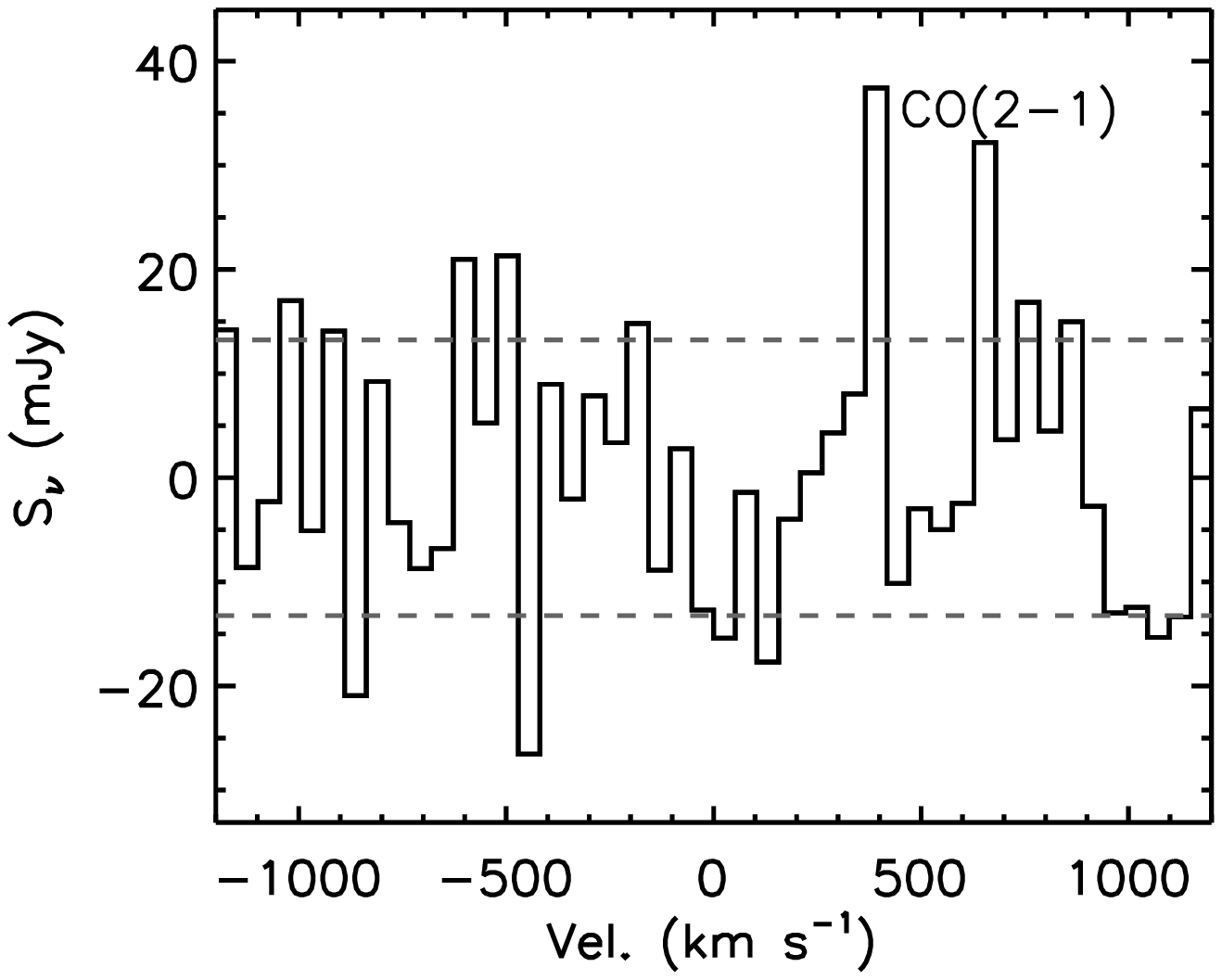}\\
\begin{turn}{90}\large \hspace{1.1cm} NGC2258\end{turn} &
\includegraphics[width=5cm,angle=0,clip,trim=0.0cm 1.3cm 0cm 0.0cm]{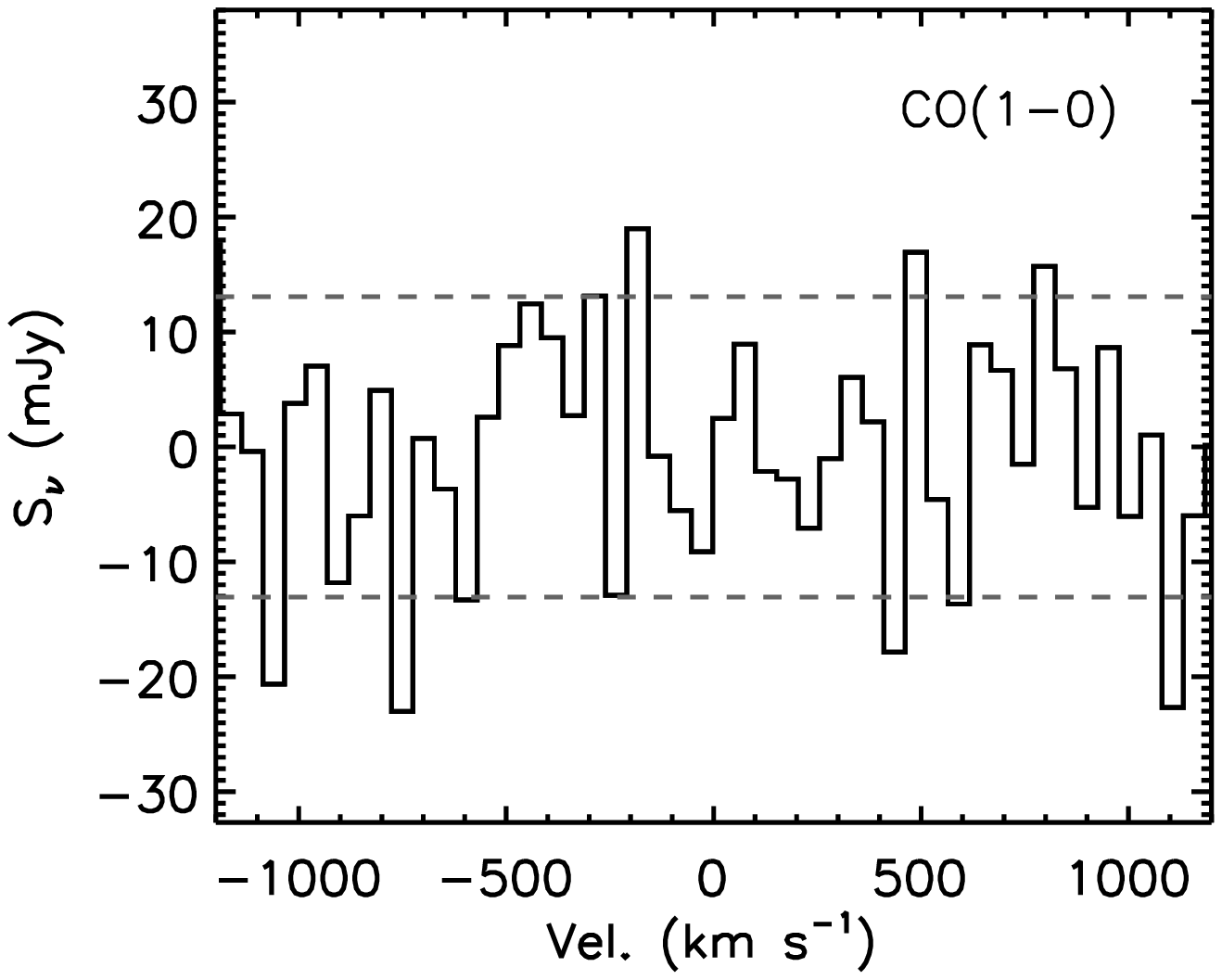} &
\includegraphics[width=5cm,angle=0,clip,trim=0.0cm 1.3cm 0cm 0.0cm]{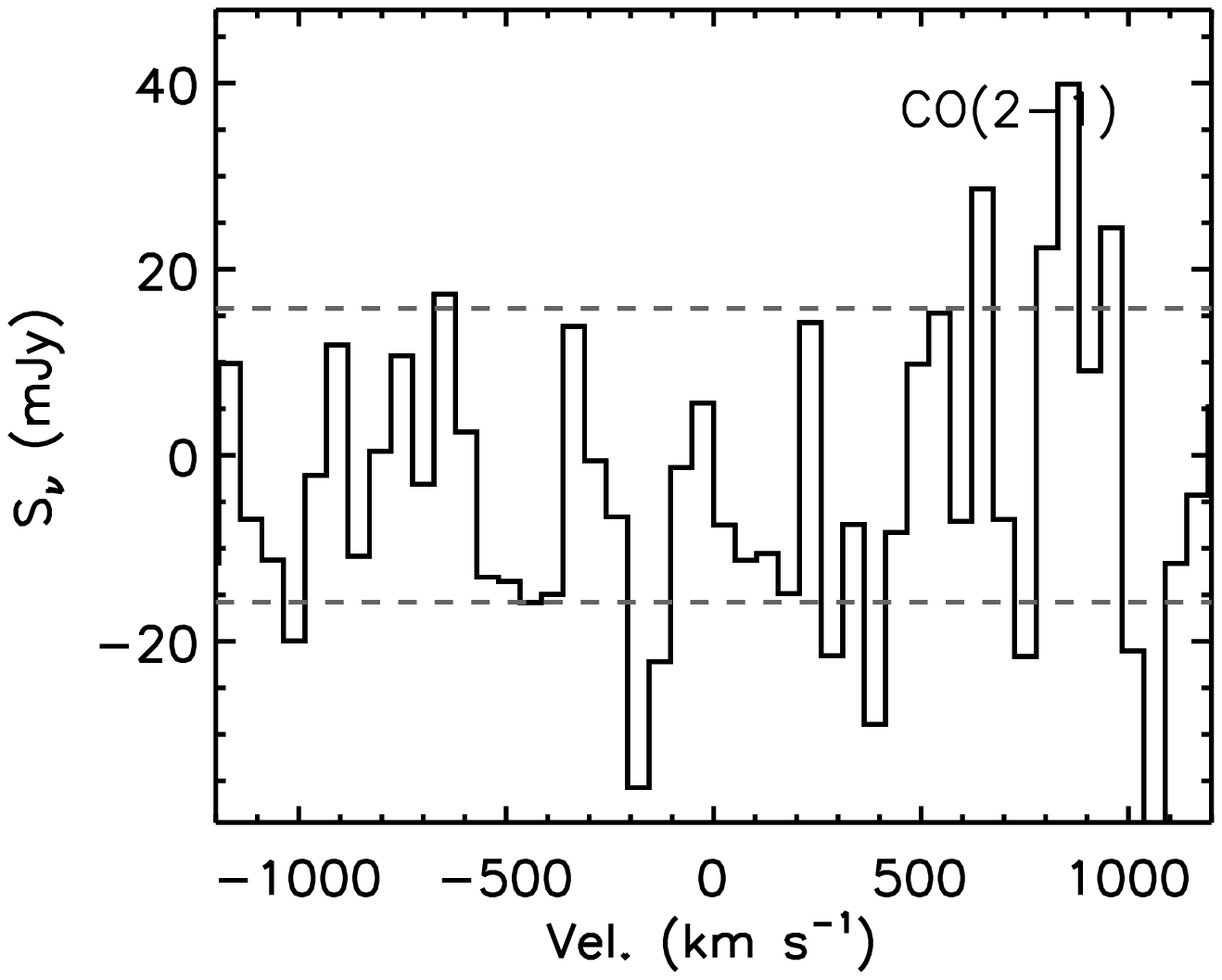} \\
\begin{turn}{90}\large \hspace{1.1cm} NGC5252\end{turn} &
\includegraphics[width=5cm,angle=0,clip,trim=0.0cm 1.3cm 0cm 0.0cm]{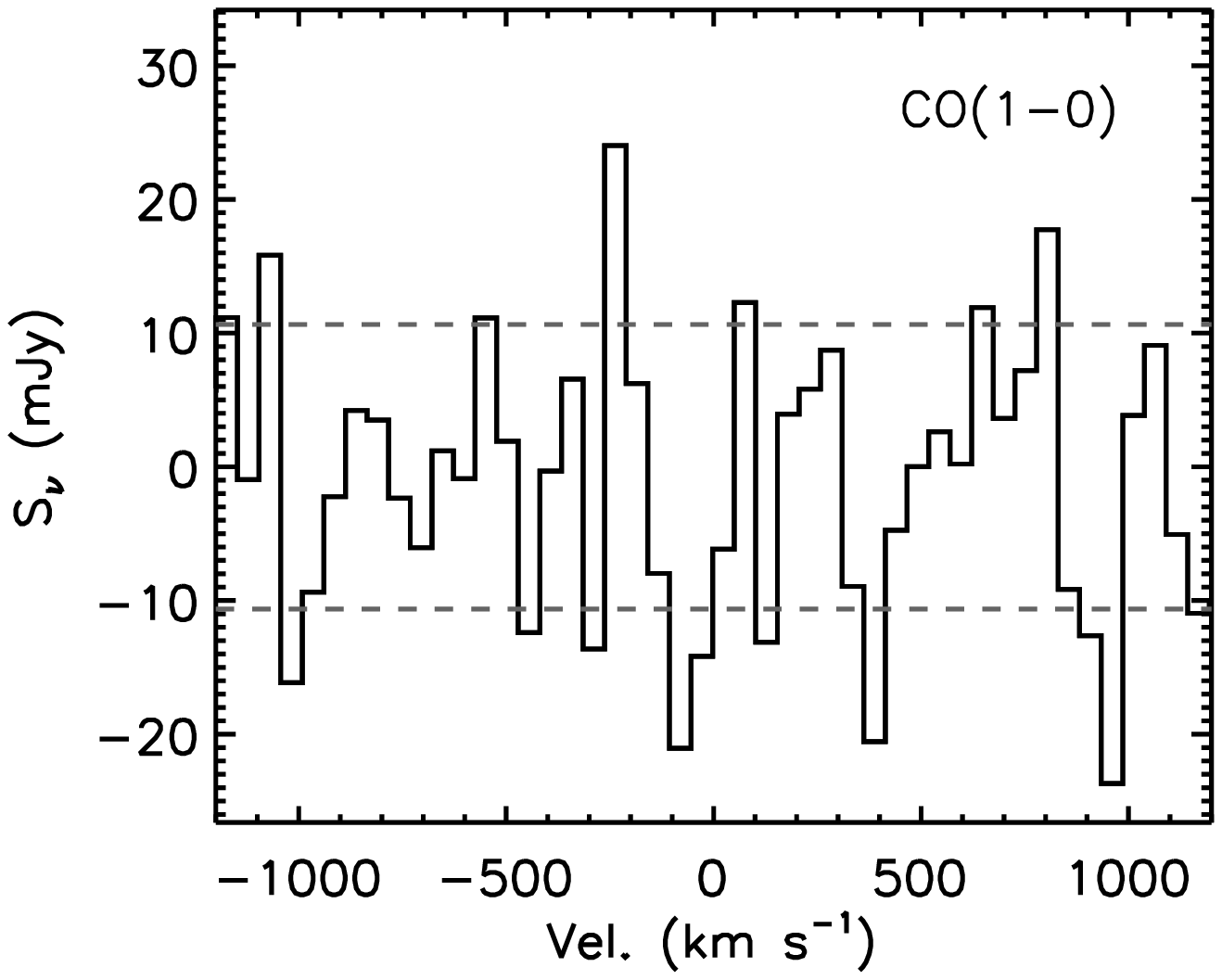} &
\includegraphics[width=5cm,angle=0,clip,trim=0.0cm 1.3cm 0cm 0.0cm]{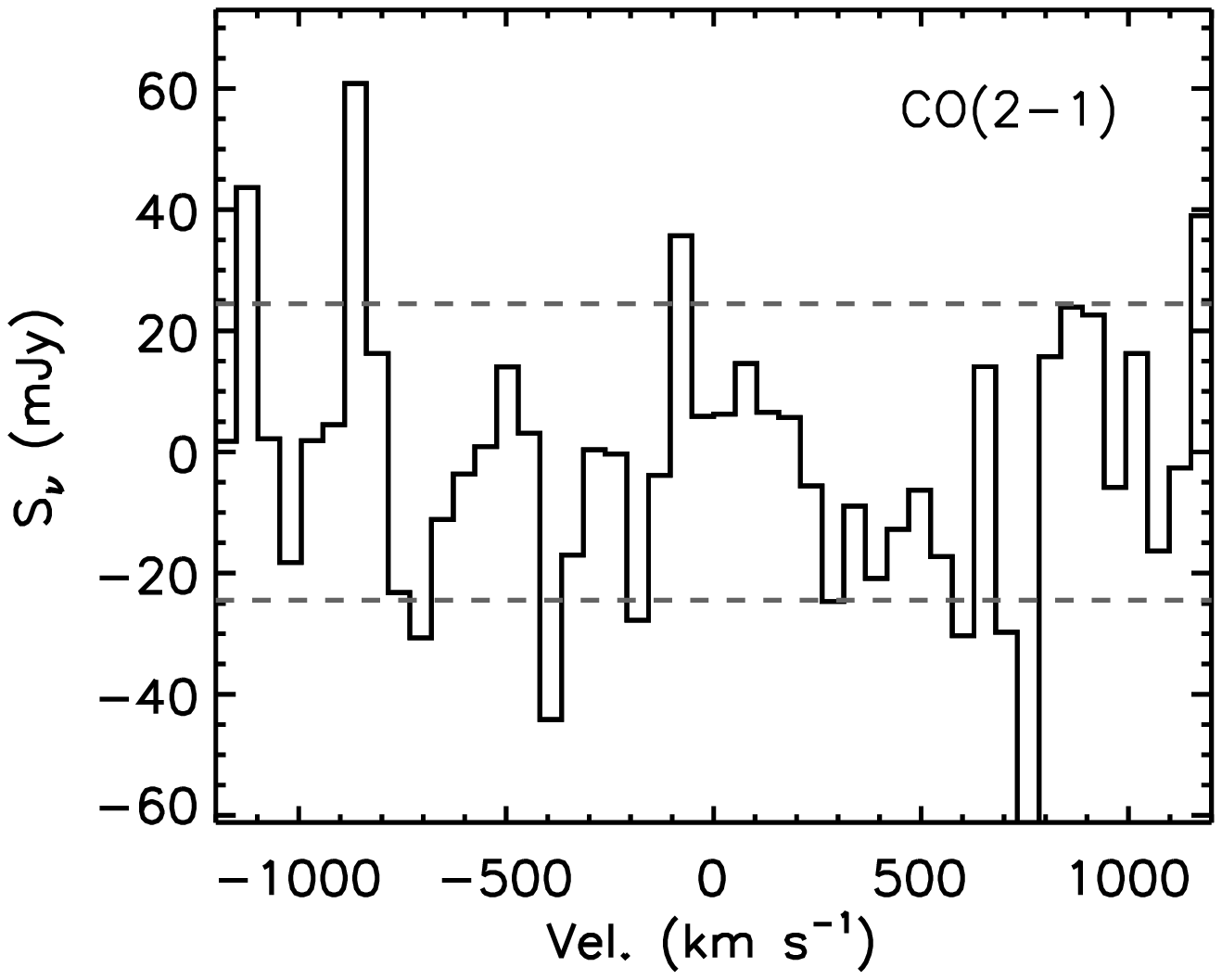} \\
\begin{turn}{90}\large \hspace{1.1cm} NGC6482 \end{turn} &
\includegraphics[width=5cm,angle=0,clip,trim=0.0cm 1.3cm 0cm 0.0cm]{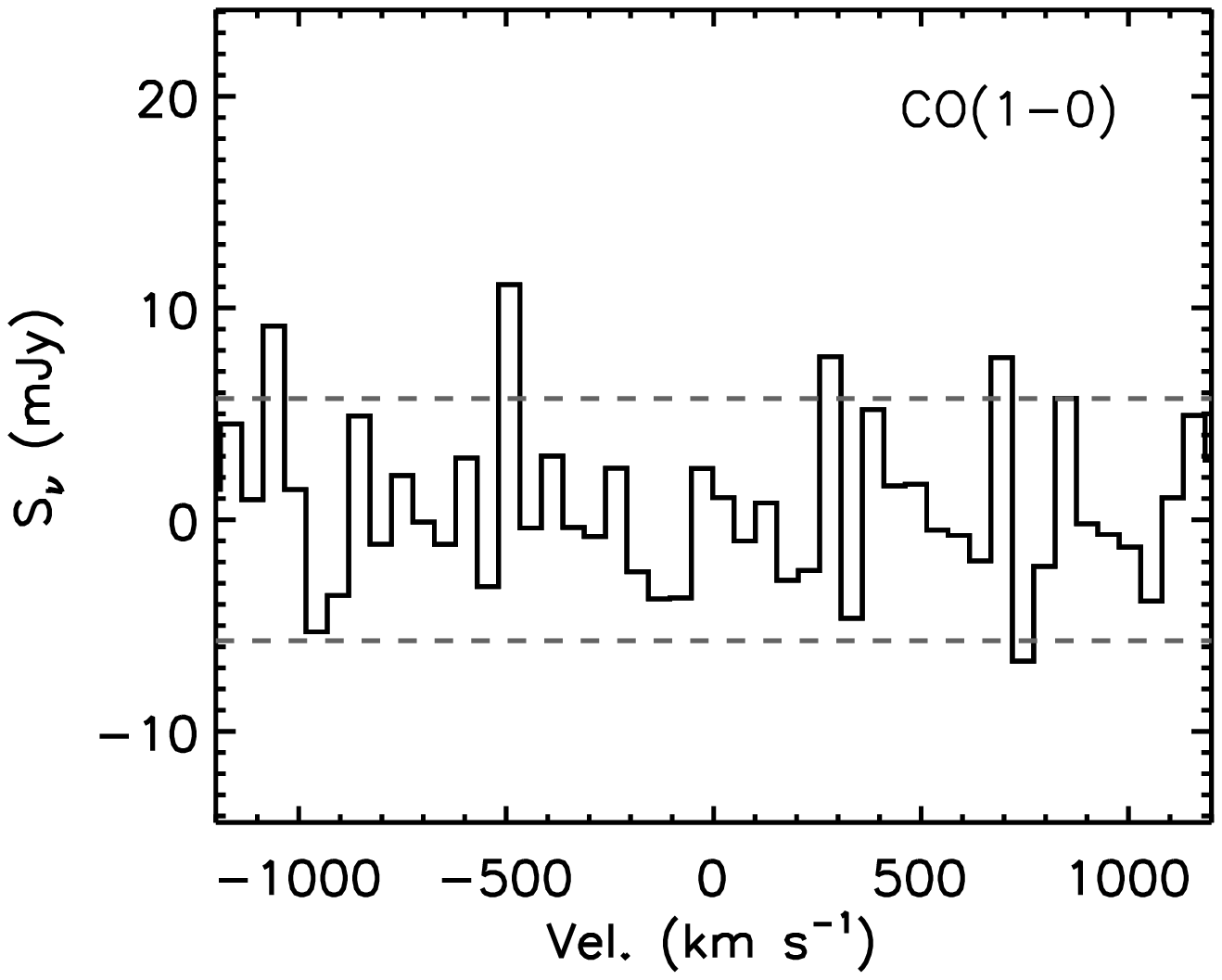} &
\includegraphics[width=5cm,angle=0,clip,trim=0.0cm 1.3cm 0cm 0.0cm]{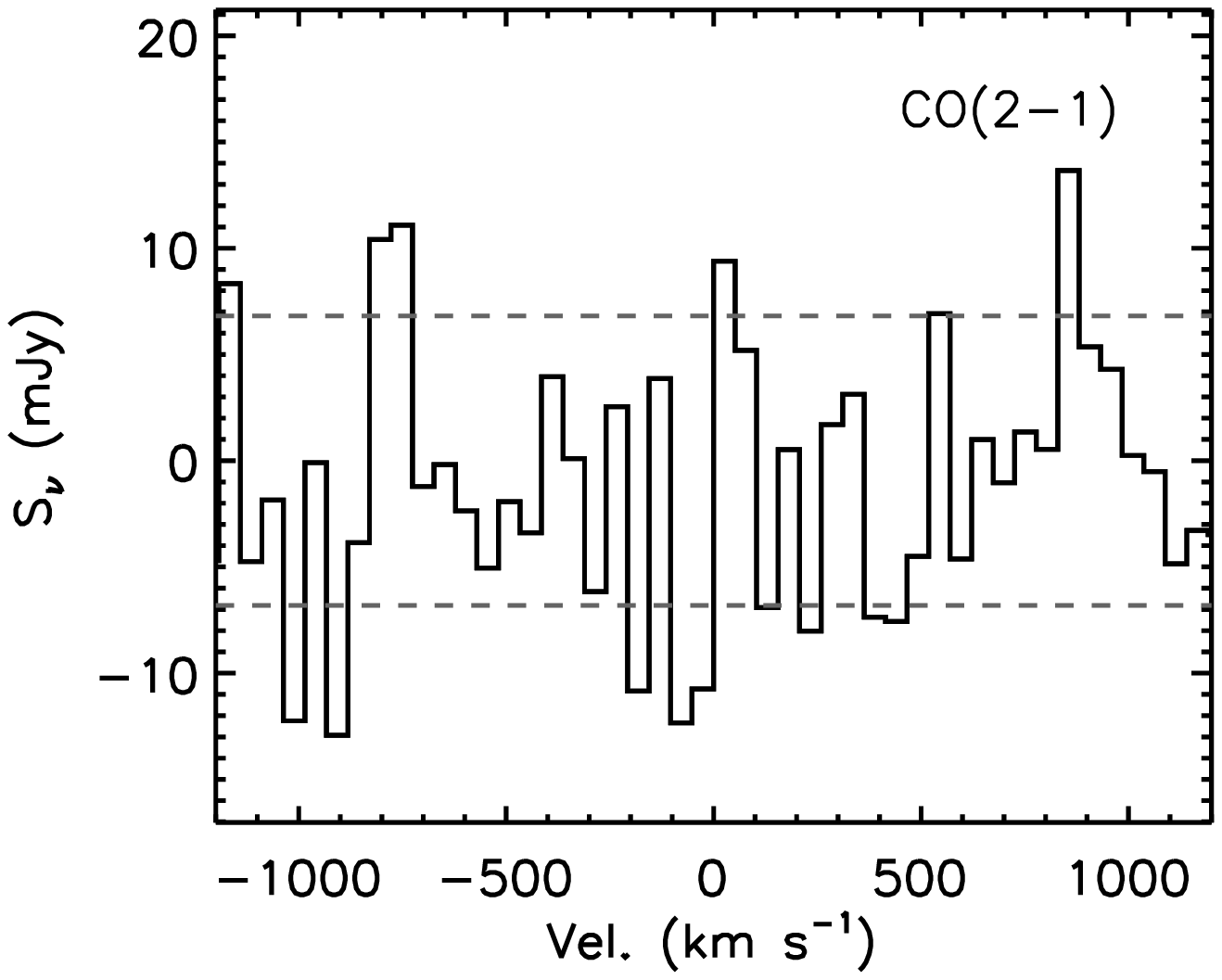}\\
\begin{turn}{90}\large \hspace{1.1cm} NGC7556 \end{turn} &
\includegraphics[width=5cm,angle=0,clip,trim=0.0cm 0.0cm 0cm 0.0cm]{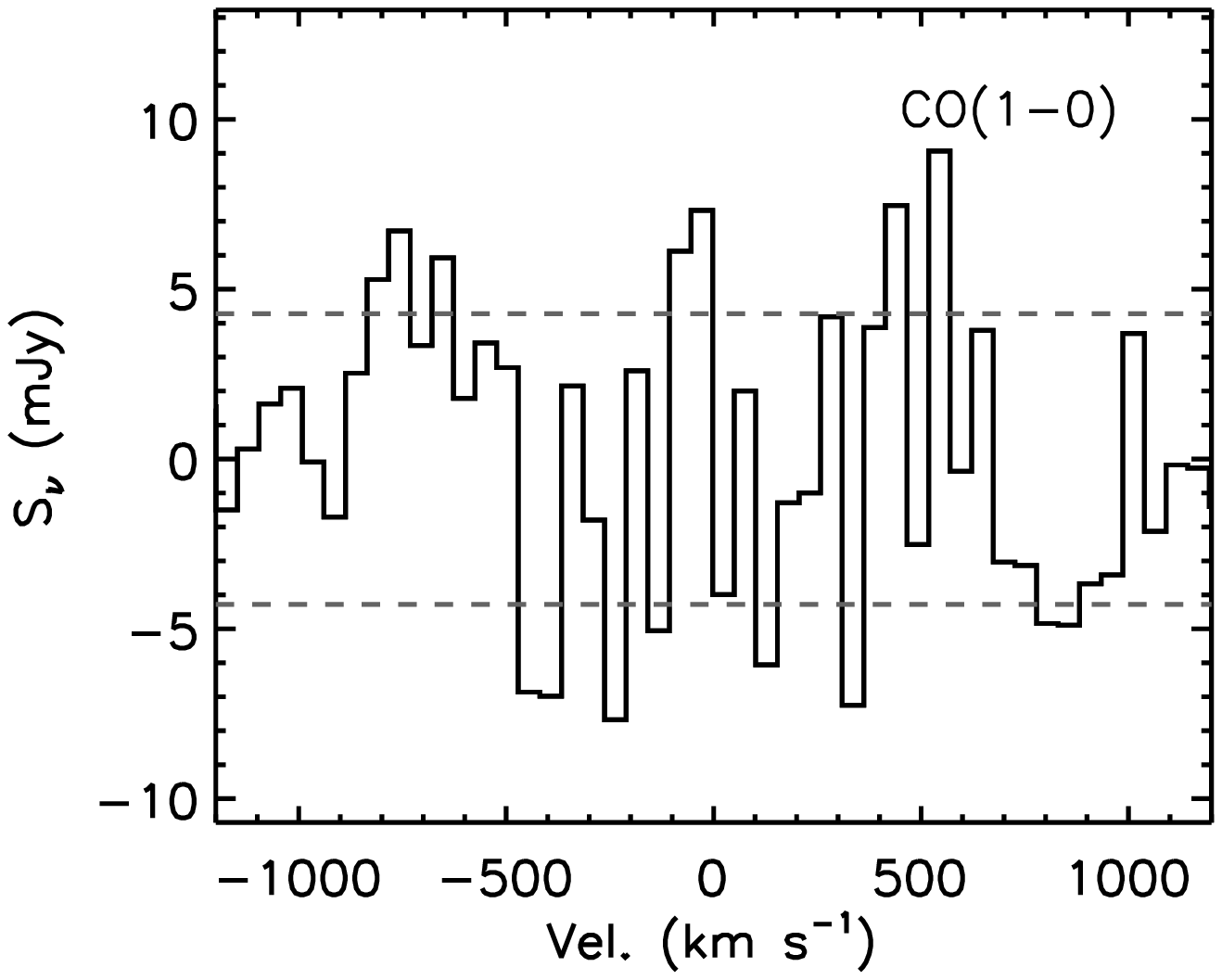} &
\includegraphics[width=5cm,angle=0,clip,trim=0.0cm 0.0cm 0cm 0.0cm]{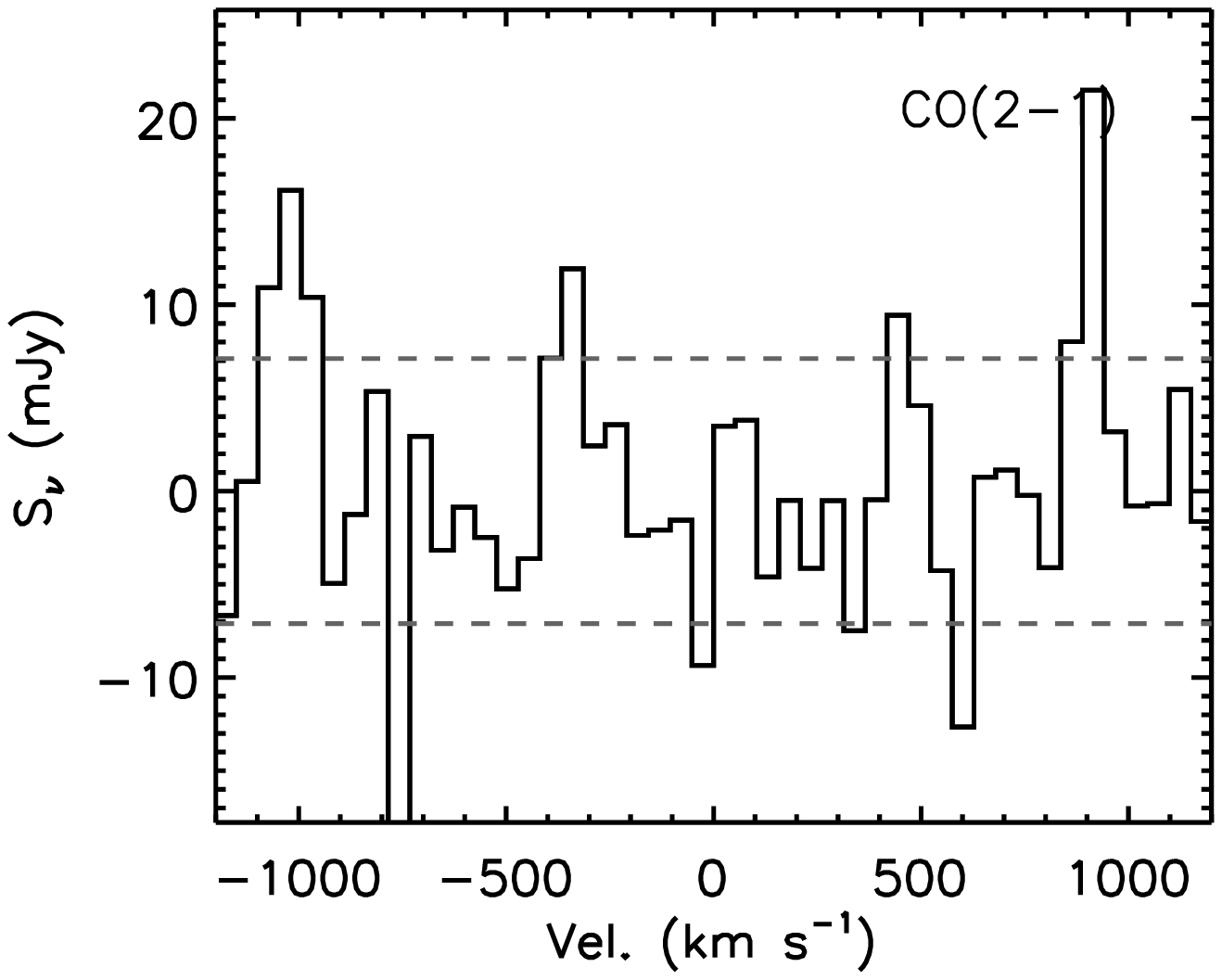}\\
\end{array}$
 \end{center}
 \caption{CO(1-0) and CO(2-1) IRAM-30m spectra (left and middle column, respectively) for our non-detected sample galaxies. The velocity on the x-axis is plotted over $\pm$1200\kms\ with respect to the rest-frequency of the line at the redshift given in Table \ref{proptable}. The dashed lines show the $\pm$1$\sigma$ RMS level.}
 \label{conodetsfig}
 \end{minipage}
 \end{figure*}

\end{document}